\documentclass[12pt]{iopart}

\usepackage{amssymb,amsfonts,bm}
\usepackage{setstack}
\usepackage{mathrsfs}
\usepackage{color,psfrag}
\usepackage{iopams}
\usepackage{stackrel}

\usepackage{graphics,epsfig,subcaption}
\usepackage{amssymb,amsfonts,bm}
\usepackage{mathrsfs}
\usepackage{color,psfrag}
\usepackage{iopams}
\usepackage{stackrel}

\newcommand{\nn}{\nonumber}
\newcommand{\bb}{\begin{eqnarray}}
\newcommand{\ee}{\end{eqnarray}}

\renewcommand{\eref}[1]{(\ref{#1})}
\renewcommand{\Re}{{\rm Re}}

\renewcommand{\e}{e}

\newcommand{\ff}{\frac{1}{2}}

\newcommand{\drm}{{\rm d}}
\renewcommand{\fl}{\hspace{-1cm}}
\newcommand{\com}{\textcolor{black}}

\newcommand{\ar}{a_{{\mathbf r}}}
\newcommand{\abr}{\bar a_{{\mathbf r}}}

\newcommand{\arx}{a_{\mathbf{r-x}}}
\newcommand{\abry}{\bar a_{\mathbf{r-y}}}

\newcommand{\Br}{{\mathbf{r}}}

\newcommand{\Bx}{{\mathbf{x}}}
\newcommand{\By}{{\mathbf{y}}}
\newcommand{\Bk}{{\mathbf{k}}}

\newcommand{\Bd}{{\mathbf{d}}}

\newcommand{\ak}{a_{{\mathbf k}}}
\newcommand{\abk}{\bar a_{{\mathbf k}}}
\newcommand{\amk}{a_{\mathbf{-k}}}
\newcommand{\abmk}{\bar a_{\mathbf{-k}}}

\newcommand{\ftwo}{\makebox{\small $\frac{1}{2}$}}

\newcommand{\HSO}{H_{S}}
\newcommand{\HS}{H_{SP}}

\newcommand{\ZSO}{Z_{S}}
\newcommand{\ZS}{Z_{SP}}

\renewcommand{\NP}{N_P}
\newcommand{\NS}{N_S}
\newcommand{\NSN}{N}

\newcommand{\ZSP}{Z_{SP}}

\newcommand{\Zk}{Z_{\Bk}(z)}
\newcommand{\FZ}{\Phi}
\newcommand{\CC}{G}
\newcommand{\rhop}{\rho}
\newcommand{\Sr}{{\mathscr{S}}_{{\mathbf r}}}
\newcommand{\SO}{{\mathscr{S}}_S}
\newcommand{\SP}{{\mathscr{S}}_{SP}}

\newcommand{\Sint}{{\mathscr{S}}_{int}}
\newcommand{\uu}{\beta_0}
\newcommand{\uux}{\beta_x}
\newcommand{\uuy}{\beta_y}
\newcommand{\uuxy}{\beta_{xy}}
\newcommand{\vx}{\alpha_x}
\newcommand{\vy}{\alpha_y}
\newcommand{\bO}{{\mathbf{0}}}
\newcommand{\ca}{\alpha_0}
\newcommand{\cb}{\beta_0}
\newcommand{\mO}{{m_{0}}}
\newcommand{\Lc}{L_c}

\newcommand{\al}{a_1}
\newcommand{\be}{a_2}

\newcommand{\arctanh}{\mathrm{arctanh}}

\begin{document}
\title[Crystallization and dynamics of defects in a magnetic fluctuating medium]{Crystallization and dynamics of defects in a magnetic fluctuating medium}
\author{Jean-Yves Fortin}
\address
{\com{Laboratoire de Physique et Chimie Th\'eoriques,
CNRS UMR 7019, 
\\ Universit\'e de Lorraine,
F-54000 Nancy, France}
}
\ead{jean-yves.fortin@univ-lorraine.fr}
\begin{abstract}{We consider the dynamics of classical particles or defects moving in a fluctuating two-dimensional magnetic medium made of Ising spins. These defects occupy empty sites, and each of them can move according to simple rules, by exchanging its location with one of the neighboring or distant spin if the energy is favorable, conserving the magnetization. We use a fermionic representation of the theory in order to map the partition function into an integral over Grassmannian variables. This model of annealed disorder can be described by a Grassmannian action containing quartic interaction terms. We study the critical behavior of this system as well as
the entropy, specific heat, and residual correlation functions which are evaluated within this Grassmannian formalism. We found in particular that the correlations are strongly
attractive at short distances in the low temperature regime and for a broader range of distances near the spin critical regime, and slightly repulsive at large distances. These results are compared with Monte-Carlo simulations.}
\end{abstract}
\section{Introduction}
The emergence of cooperative phenomena in interacting spin systems, quantum or classical, with mobile defects \cite{Scesney:1970} reveals the role of the magnetic fluctuations in the dynamics of these defects. For example there is evidence of the existence of correlated stripes of charges or holes in two-dimensional copper oxide superconducting materials that are closely related to the commensurability between the lattice and the spin spatial structure when dilution is present \cite{Tranquada:1995}. In the classical case of dilute 
spin models where annealing disorder is represented by moving holes or defects, critical and thermodynamical properties have been investigated using exact models on special lattice structures.
In early works, exact calculations were performed on decorated square, hexagonal, or diced lattices \cite{Syozi:1966} where magnetic or nonmagnetic ions occupy a sublattice formed by the centers of the bonds between magnetic sites on the main lattice. The magnetic ions of the main lattice interact via couplings which can be ferromagnetic, antiferromagnetic, or zero depending on the presence of magnetic or nonmagnetic ions on the bonds. When the concentration of magnetic ions on the sublattice is varied,
there is a specific value below which magnetic order vanishes and cluster percolation disappears. Annealed Ising models with site dilution on self-similar or hierarchical structures were also studied, for example on the triangular Apollonian network \cite{Andrade:2005,Silva:2014} with a random distribution of vacant sites that are thermalized. The transfer matrix method in the Grand Ensemble reveals that the presence of magnetic order depends not only on the chemical potential of vacancies but also on the lattice choice. 
In these models it is also worth noticing the presence of a residual macroscopic entropy at zero temperature due to the complex geometrical and random connectivity between magnetic sites, and the contribution of nonactive magnetic sites. Non zero residual entropy is also seen in spin-ice models with quenched disorder on similar hierarchical lattices mimicking the tetrahedral three dimensional structures \cite{Fortin:2013}. However the precise value of the residual entropy is not always understood in general in terms of simple local spin configurations contributing to highly degenerate ground states.
Dynamical properties of mobile defects were also studied on a dilute Ising chain and within the Bethe approximation \cite{Semkin:2016,Semkin:2019,Semkin:2022}, when a short range potential is introduced between magnetic ions. Spatial correlation functions were evaluated between the mobile impurities as function of the dilution and different parameters such as the temperature or potential strength, which reveals the degree of randomness of the impurities in the system at thermal equilibrium.

In this context, we would like to analyze the degree of correlation between mobile defects in the
simple case of the classical dilute Ising model on a square lattice. In particular, the knowledge of the spatial correlation functions between mobile defects at equilibrium can shed light on the contribution of the local magnetic fluctuations to the dynamics of these impurities. We expect that these magnetic fluctuations induce some
effective potential between the non-magnetic particles, which can be seen by studying the different 
thermodynamical quantities as well. Analytically, we will consider a method based on Grassmann variables in order to fermionize the theory, in parallel to a previous work on quenched disorder \cite{Plechko:98}. We will show that we can transform the original spin and defect or hole model into a non-linear fermionic action, depending on the positional configuration of the defects represented by bosonic nilpotent variables. The advantage of this formulation is that corrections to non-linear terms are usually small due to the nilpotent properties of the Grassmann variables and good approximations to the thermodynamical quantities can be achieved. The present model is actually similar to a three-states Blume-Capel model \cite{Clusel:2008}, with spin zero representing the defects, however the main difference is that the number of variables with spin zero representing the presence onsite of nonmagnetic impurities is fixed by their concentration, either in the Canonical Ensemble or in the Grand Canonical Ensemble with a chemical potential.

From a numerical point of view, we have performed Monte-Carlo (MC) simulations and investigated two cases for the dynamics of the mobile defects, considered as hard core particles: first we will study the local diffusion (model A), where a particle can hop to one of the four neighboring sites by exchange of
the local state with the corresponding spin, if the energy is favorable and if no other particle is already present. In the second case (model B), we consider that particles can hop to any site distant on the lattice with the same previous rules. In both cases there is no change in the magnetization. Configurations with the highest probability are counted by the partition function mixing particles and spins. In particular, we observe that the low temperature properties display 
a crystallization of small clusters of particles in the model A, due to the fact that it is favorable
energetically for particles to aggregate in a medium where most of the spins are polarized in one
direction. Once a particle reaches a local cluster, its motion is frozen since there is almost
no possibility that it can leave this cluster. The number of small clusters depends on the space configuration of the particles just above the crystallization temperature.
In the model B, we expect that all particles aggregate into a single giant cluster due to the 
possibility of reducing the total surface of the clusters by non-local diffusion.

\begin{figure*}[h!]
\centering
\begin{subfigure}[b]{0.4\linewidth}
\caption{}
\includegraphics[clip]{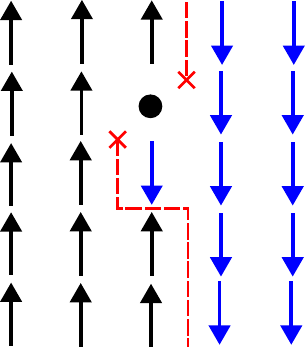}
\end{subfigure}
\begin{subfigure}[b]{0.4\linewidth}
\caption{}
\includegraphics[clip]{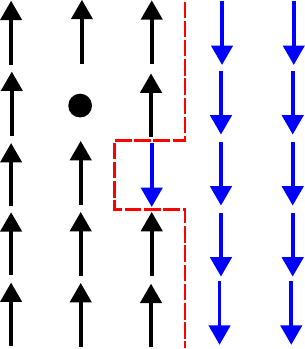}
\end{subfigure}
\caption{
Spin configurations between two domains of opposite polarity, with a particle (solid circle) along 
the domain wall, case (a), or inside the domain of spins up, case (b). The energy for the spin
configuration (a) has a lower energy than (b), and the energy difference is $4J$.
The red crosses in (a) indicate that the domain wall is interrupted by the presence of the particle.
}\label{fig_spins_hole}
\end{figure*}

In the high temperature regime however, the particles tend to localize near the domain walls, as seen
in the two examples of figure \ref{fig_spins_hole}, since it is energetically beneficial to take advantage of the discontinuity in spin polarity, especially near the critical regime. The interparticle correlations can be evaluated through the thermal space correlation functions. At high temperature, we expect that the strong thermal fluctuations will not contribute to the particle dynamics, and that free diffusion of uncorrelated particles will occur.

The paper is organized as follow: In section \ref{sect_a} we introduce the model and explain
how to transform the Hamiltonian into a fermionic action, using Grassmann algebra. This is an 
extension of the Grassmann representation of the Ising model that includes defects. In relation with
fluid theory, we introduce the notion of fugacity or activity in the description of our model.
In section \ref{sect_b}, we analyze the critical properties of the model, in particular we give an
estimation of the critical concentration of defects above which magnetization disappears.
In section \ref{sect_c}, we consider the entropy and specific heat and discuss the differences between
the analytical results and simulations in the low temperature regime. This is supplemented by the
computation of the two-point correlation functions in section \ref{sect_d} where an estimation of the
long distance behavior is given. Finally we summarize and discuss the physical findings of this work in the conclusion, section \ref{sect_e}. 

\section{Effective interaction with the magnetic fluctuations \label{sect_a}}
%
In this part, we introduce the model and consider a square lattice of size $N=L^2$ on which each site is occupied either by 
a spin or a non-magnetic particle or vacancy. The spin Hamiltonian describing this system is represented by an Ising model with ferromagnetic coupling $J>0$
\bb\label{Ham}
\HS=-\frac{J}{2}\sum_{<\Br,\Br'>}(1-n_{\Br})\sigma_{\Br}(1-n_{\Br'})\sigma_{\Br'},
\; \sigma_{\Br}=\pm 1,\; n_{\Br}=0,1,
\ee
where the sites occupied by diffusive particles are represented by the variable $n_{\Br}=1$,
whereas the spins are represented by the local state $(\sigma_{\Br}=\pm 1,n_{\Br}=0)$. The total number of vacancies on the lattice is given by $\NP\le N$.
The brackets $<\Br,\Br'>$ take into account only next neighboring sites $\Br$ and $\Br'$.
The subscripts $S$ and $P$ indicate, in the subsequent notations, the spins and particles respectively. It is well known that the Ising spin system in absence of vacancies, or $\HSO=-\ftwo J\sum_{<\Br,\Br'>}\sigma_{\Br}\sigma_{\Br'}$, has a fermionic representation of its partition function $\ZSO$ in terms of one pair of Grassmann variables $(\ar,\abr)$ per site \cite{Plechko:96p,Plechko:97}
\bb
\ZSO=
\prod_{\Br}\sum_{\sigma_{\Br}=\pm 1} \e^{-\beta \HSO}=
2^N\cosh(\beta J)^{2N}\int \left (\prod_{\Br}\drm \abr\drm \ar\right ) \e^{\SO},
\ee
where the Grassmannian action $\SO$ is defined by $\SO=\sum_{\Br}\Sr$ with the local quantities
\bb
\label{SIsing}
\Sr=\ar\abr+u(\ar+\abr)(\arx-\abry)-u^2\arx\abry.
\ee
In this expression $u=\tanh(\beta J)$, and $\mathbf x$ and $\mathbf y$ are the elementary unit steps in the $x$ and $y$ directions respectively. We recall the main following Grassmannian integration rules:
$\int da_{\Br}a_{\Br}=1$, $\int da_{\Br}=0$, and $a_{\Br}\bar a_{\Br}=-\bar a_{\Br}a_{\Br}$.
The factor $2^N$ counts the number of spin configurations and one $\cosh(\beta J)$ has been factorized for each coupling in the two directions.
%
\subsection{Fermionic representation in presence of vacancies\label{sect_aa}}
%
In presence of vacancies, we can derive a similar and more general result in more details. We first consider the decoupling of the spins using two pairs of Grassmann variables per site $(\ar,\abr)$ and $(b_{\Br},\bar b_{\Br})$, or one pair for each direction, such that
\bb\nn
\ZS[\{n_{\Br}\}]
&=\prod_{\Br}\sum_{\sigma_{\Br}=\pm 1} \e^{-\beta \HS}
\\ \nn
&=\cosh(\beta J)^{2N}\prod_{\Br}\sum_{\sigma_{\Br}=\pm 1}
(1+u\epsilon_{\Br}\sigma_{\Br}\epsilon_{\Br+\Bx}\sigma_{\Br+\Bx})
(1+u\epsilon_{\Br}\sigma_{\Br}\epsilon_{\Br+\By}\sigma_{\Br+\By})
\\ \nn
&=\cosh(\beta J)^{2N}\prod_{\Br}\sum_{\sigma_{\Br}=\pm 1}
\int d\bar a_{\Br} da_{\Br}\e^{a_{\Br}\bar a_{\Br}}\int d\bar b_{\Br} db_{\Br} 
\e^{b_{\Br}\bar b_{\Br}}
\\ \label{eq_int}
&\times
(1+a_{\Br}\epsilon_{\Br}\sigma_{\Br})(1+u\bar a_{\Br}\epsilon_{\Br+\Bx}\sigma_{\Br+\Bx})
(1+b_{\Br}\epsilon_{\Br}\sigma_{\Br})(1+u\bar b_{\Br}\epsilon_{\Br+\By}\sigma_{\Br+\By}),
\ee
where we have defined $\epsilon_{\Br}=1-n_{\Br}$ for simplification. The last line can be rewritten as the product of four non-commutative quantities 
$A_{\Br}\bar A_{\Br+\Bx}B_{\Br}\bar B_{\Br+\Bx}$, which then have to be rearranged inside
the overall integral in order to group together the weights containing the same spin $\sigma_{\Br}$,
or index $\Br$, that can be summed up independently. The overall product in \eref{eq_int} can be rearranged in the correct order by considering a mirror-ordering factorization procedure, which has a valid meaning
under the sign of integration. Free boundary conditions are taken here, as it simplifies the 
problem of ordering at the border of the lattice. When double periodic boundary conditions are considered, we can cite for more details the Appendix of the reference \cite{Plechko:1985}, which proposes a straightforward solution for this problem of periodicity at the borders using a general symmetry identity for Grassmannian polynomial functions. The exponential terms in \eref{eq_int} are commuting
variables and can be factorized easily. We obtain, using the notation $\Br=(m,n)$
\bb\nn
\ZS[\{n_{\Br}\}]
&=\cosh(\beta J)^{2N}
\prod_{\Br}\sum_{\sigma_{\Br}=\pm 1}
\int d\bar a_{\Br} da_{\Br}\e^{a_{\Br}\bar a_{\Br}}\int d\bar b_{\Br} db_{\Br} 
\e^{b_{\Br}\bar b_{\Br}}
\\
&\times \overrightarrow{\prod_{n=1}^{L}}\bar B_{1n}A_{1n}\left (
\overrightarrow{\prod_{m=2}^{L}}\bar A_{mn}\bar B_{mn}A_{mn}
\overleftarrow{\prod_{m=2}^{L}}B_{mn}\right )B_{1n},
\ee
where the arrows indicate the ordering of the products. We can then perform iteratively the summation over the spin variables, since in this form the four weights that share the same spin variable
are grouped together. In particular we can exponentiate the resulting quadratic and quartic Grassmannian
polynomial terms after summation over the spin variable, which commute with the rest of the products and therefore can be factorized before the next iterative summation is performed
\bb
\sum_{\sigma_{\Br}=\pm 1}\bar A_{\Br}\bar B_{\Br}A_{\Br}B_{\Br}
=2\exp\left [a_{\Br}b_{\Br}\epsilon_{\Br}+u(\bar a_{\Br-\Bx}+\bar b_{\Br-\By})(a_{\Br}+b_{\Br})
\epsilon_{\Br}+u^2\bar a_{\Br-\Bx}\bar b_{\Br-\By}\epsilon_{\Br}\right ].
\ee
We finally obtain
\bb
\ZS[\{n_{\Br}\}]
&=2^N\cosh(\beta J)^{2N}\int \prod_{\Br}
d\bar a_{\Br} da_{\Br}d\bar b_{\Br} db_{\Br} 
\\ \nn
&\times
\exp\left [\sum_{\Br} a_{\Br}\bar a_{\Br}+b_{\Br}\bar b_{\Br}
+a_{\Br}b_{\Br}\epsilon_{\Br}+u(\bar a_{\Br-\Bx}+\bar b_{\Br-\By})(a_{\Br}+b_{\Br})
\epsilon_{\Br}+u^2\bar a_{\Br-\Bx}\bar b_{\Br-\By}\epsilon_{\Br}\right ].
\ee
We can further reduce the number of variables by performing an integration over $(a_{\Br},b_{\Br})$, which leads to a simplification of the previous expression
\bb
\ZS[\{n_{\Br}\}]
&=2^N\cosh(\beta J)^{2N}\int \prod_{\Br}
d\bar a_{\Br}d\bar b_{\Br} 
\\ \nn
&\times
\exp\left [\sum_{\Br} \bar b_{\Br}\bar a_{\Br}
+u(\bar a_{\Br-\Bx}+\bar b_{\Br-\By})(\bar b_{\Br}-\bar a_{\Br})
\epsilon_{\Br}+u^2\bar a_{\Br-\Bx}\bar b_{\Br-\By}\epsilon_{\Br}\right ].
\ee
This expression is compared to \eref{SIsing} in the case where $\epsilon_{\Br}=1$, 
by formally replacing  the pair $(\bar a_{\Br},\bar b_{\Br})
\rightarrow (a_{\Br},-\bar a_{\Br})$, which leads to the expression of $\Sr$ containing only
one pair of Grassmann variables per site. In the general case, we obtain instead \cite{Plechko:98,Plechko:99p,Plechko:10}
\bb
\ZS[\{n_{\Br}\}]
&=2^N\cosh(\beta J)^{2N}\int \prod_{\Br}
d\bar a_{\Br}da_{\Br} 
\\ \nn
&\times
\exp\left [\sum_{\Br} a_{\Br}\bar a_{\Br}
+u(a_{\Br}+\bar a_{\Br})(a_{\Br-\Bx}-\bar a_{\Br-\By})
\epsilon_{\Br}-u^2a_{\Br-\Bx}\bar a_{\Br-\By}\epsilon_{\Br}\right ].
\ee
In terms of the variables $\{n_{\Br}\}$ describing the particle positions, the partition function can be expressed as
\bb
\ZS[\{n_{\Br}\}]
=2^{\NS}\cosh(\beta J)^{2\NS}\int \prod_{\Br}\drm \abr\drm \ar 
\left [ \delta_{n_{\Br},0}\e^{\Sr}+\delta_{n_{\Br},1}\ar\abr \right ],
\ee
where we have replaced the number of spins $N$ in the overall factor by the number of active spins
$\NS=N-\NP$, since there are $\NP$ hidden spins which does not contribute to the thermodynamics. Indeed we have to formally remove the spin degrees of freedom per site occupied by the particles as well as the absence of couplings between spins and particles.
This partition function can be further rewritten as
\bb\label{ZS}
\ZS[\{n_{\Br}\}]
=2^{\NS}\cosh(\beta J)^{2\NS}\int \prod_{\Br}\drm \abr\drm \ar \left [ 1+\left (\ar\abr\e^{-\Sr}-1\right )n_{\Br} \right ]\e^{\SO}.
\ee
To simplify the problem, we first diagonalize $\SO$, using the Fourier space transformation $\ar=L^{-1}\sum_{\Bk}\ak\e^{i\Bk.\Br}$, $\abr=L^{-1}\sum_{\Bk}\abk\e^{-i\Bk.\Br}$, and express the partition function $\ZSO$ as a product over momentum dependent factors, after relabeling momenta $k_x \rightarrow -k_x$:
\bb\nn
\SO&=&
{\sum_{\Bk}}^{'}
\left [g_{\Bk}\ak\abk+\bar g_{\Bk}\amk\abmk-2iu\sin(k_x)\ak\amk
+2iu\sin(k_y)\abk\abmk\right ],
\\ \label{S0k}
g_{\Bk}&=&1-u(\e^{ik_x}+\e^{ik_y})-u^2\e^{ik_x+ik_y}.
\ee
The partition function $\ZSO$ can then be easily factorized using Grassmannian integral rules $(\NS=N)$
\bb\nn
\ZSO&=&2^{N}\cosh(\beta J)^{2N}\int 
{\prod_{\Bk}}^{'}
d\abk d\ak d\abmk d\amk
\e^{\SO}
\\ \nn
&=&2^{N}\cosh(\beta J)^{2N}{\prod_{\Bk}}^{'}
\left (g_{\Bk}\bar g_{\Bk}-4u^2\sin(k_x)\sin(k_y) \right )
\\ 
&=&2^{N}\cosh(\beta J)^{N}{\prod_{\Bk}}^{'}
\left [ (1+u^2)^2-2u(1-u^2)(\cos(k_x)+\cos(k_y))\right ].
\ee
%
%
\begin{figure}[!hb]
\centering
\includegraphics[angle=0,scale=1,clip]{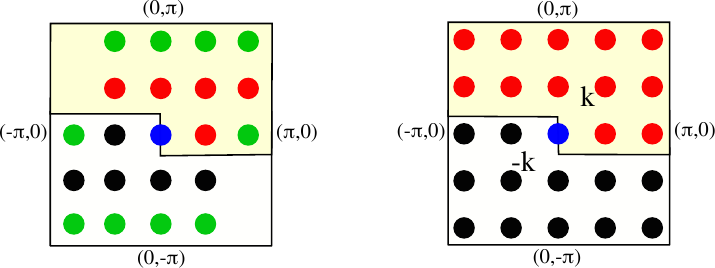}
\caption{
Reduced Brillouin zone located in the yellow shade area with momenta $\Bk$, or red circles,
contributing to the pairs of Grassmann variables $(\abk,\ak)$, for $L$ even $L=4$ (left) and odd $L=5$
(right). The other momenta, or black circles, are related by the relation $\Bk\rightarrow -\Bk$
and contribute to the pairs $(\abmk,\amk)$. The blue circle is the zero momentum $\Bk=0$ corresponding to $(\bar a_{{0}},a_{{0}})$. When $L$ is even, momenta in green have their counterparts in both zones
since $a_{k_x,L/2}=a_{k_x,-L/2}$ or $a_{L/2,k_y}=a_{-L/2,k_y}$, and should be counted with a factor $1/2$.}\label{fig_Brillouin}
\end{figure}
%
The prime symbol corresponds to the set of momenta $\Bk$ located in one half of the Brillouin zone, the opposite momenta $''-\Bk''$ completing the zone, see figure \ref{fig_Brillouin}. The opposite momentum
$''-\Bk''$ is related to $\Bk$ by the fact that $(\Bk+''-\Bk'')$ is zero modulo the period $(2\pi,2\pi)$ of the lattice.  
The particular momentum contribution $(\bar a_{{0}},a_{{0}})$ could be included formally into the reduced sum 
but this term should be negligible in the thermodynamical limit as it corresponds to a single value in the large asymptotic sum and will be omitted.
One can write an effective action for (\ref{ZS}) by introducing nilpotent commuting variables $\rho_{\Br}$ such that $\rho_{\Br}^2=0$ and
\bb\label{ZSpart}
\ZS[\{n_{\Br}\}]&=&2^{\NS}\cosh(\beta J)^{2\NS}
\\ \nn
&\times&
\int \prod_{\Br}\drm \abr\drm \ar\drm \rho_{\Br} \exp\left (\SO+\sum_{\Br}\rho_{\Br}(1-n_{\Br})+\sum_{\Br}\rho_{\Br}n_{\Br}\ar\abr\e^{-\Sr}\right ).
\ee
One should notice that $\rho_{\Br}$ can always be written as a product of two Grassmann variables.
The term $\ar\abr\e^{-\Sr}=\ar\abr-u^2\ar\arx\abr\abry$ will be approximated in the following by the quadratic part since the quartic part contains second order derivatives which will be assumed to be negligible for small momenta and small concentration of particles. However the corrections to the quartic part are studied in \ref{app_1}. We can now construct the total partition function $\ZSP$ in the Grand Canonical Ensemble for the spin-particles system that includes the vacancies in thermal equilibrium
\bb\nn
\ZSP&=\prod_{\Br}\sum_{n_{\Br}=0,1}\ZS[\{n_{\Br}\}]\exp\left (\beta\mu\sum_{\Br}n_{\Br}\right )
\\ \nn
&=2^{\NSN}\cosh(\beta J)^{2\NSN}\int \prod_{\Br}\drm \abr\drm \ar \exp\left (\SO+z\sum_{\Br}\ar\abr \e^{-\Sr}\right )
\\ \label{ZGC}
&=2^{\NSN}\cosh(\beta J)^{2\NSN}\int \prod_{\Br}\drm \abr\drm \ar \exp(\SP),
\ee
where we have introduced a chemical potential $\mu$ that fixes the average number of particles $\NP$
in this representation. The parameter $z=\e^{\beta\mu}/2\cosh^2(\beta J)$ is called the activity or fugacity, incorporating a spin chemical potential that we can define as being equal to $-\log(2\cosh^2(\beta J))/\beta\simeq -2J$ at low temperature. The parameter $\NS=N-\NP$ in the overall factor has been replaced by $N$ and the corrective factor,
corresponding to remaining part $\NP$, has been included with the chemical potential in the definition of $z$. The expression \eref{ZGC} is exact and comprises not only quadratic but also quartic terms in the Grassmannian action, which can not be integrated by means of a determinant without approximation. The chemical potential is defined by the implicit relation
\bb
\NP=-\frac{\partial \Omega}{\partial \mu},\;
\Omega=-k_BT\log \ZSP,
\ee
with $\Omega$ the Grand Potential. After summation over the particle variables and using the quadratic approximation for the term in factor of $z$, the effective action $\SP$ in equation \eref{ZGC} can be written in the Fourier space as
\bb\label{Sk}\fl
\SP=
{\sum_{\Bk}}^{'}
\Big [(z+g_{\Bk})\ak\abk+(z+\bar g_{\Bk})\amk\abmk-2iu\sin(k_x)\ak\amk
+2iu\sin(k_y)\abk\abmk\Big ].
\ee
When there is no particle in the system, $\mu=-\infty$ and $z=0$ as expected. It is straightforward to perform the integration over the Grassmann variables in \eref{Sk}, which leads to the Fourier factorization
\bb\nn
\ZSP=2^{\NSN}\cosh(\beta J)^{2\NSN}{\prod_{\Bk}}^{'}
\Big [(z+g_{\Bk})(z+\bar g_{\Bk})-4u^2\sin(k_x)\sin(k_y) \Big ]
\\ \nn
=2^{\NSN}\cosh(\beta J)^{2\NSN}{\prod_{\Bk}}^{'}
\Big [ (1+z)^2+u^2(2+u^2)-2u(1+z-u^2)(\cos(k_x)+\cos(k_y))
\\ \label{eq_Zk}
-2u^2z\cos(k_x+k_y)\Big ]=2^{\NSN}\cosh(\beta J)^{2\NSN}{\prod_{\Bk}}^{'}
\Zk.
\ee
We will consider in the following the free energy $F$ in the Canonical Ensemble, for a fixed concentration of defects, using the Legendre transformation 
%
\bb
F=\Omega+\mu\NP.
\ee
%
%
\subsection{Equation for the fugacity\label{sect_ab}}
%
The average number of particles is then defined by $\NP
=z\partial_z\log\ZSP$, and the density by $\rhop=\NP/N$.
The local density $n_{\Br}$ is represented formally by the Grassmann variable
$z\ar\abr\e^{-\Sr}\simeq z\ar\abr$ in the averaging process. If we define $\langle\cdots\rangle$ as the Boltzmann thermal average, then we have
\bb\nn
\langle n_{\Br}\rangle=z\langle \ar\abr \rangle&=
\frac{z}{N}{\sum_{\Bk,\Bk'}}^{'}
\Big \langle
\left (a_{\Bk} \e^{i\Bk.\Br}+a_{-\Bk} \e^{-i\Bk.\Br}\right )
\left (\bar a_{\Bk'} \e^{-i\Bk'.\Br}+\bar a_{-\Bk'} \e^{i\Bk'.\Br}\right )
\Big \rangle
\\
&=\frac{z}{N}{\sum_{\Bk}}^{'}
\Big \langle
a_{\Bk}\bar a_{\Bk}+a_{-\Bk}\bar a_{-\Bk}
\Big \rangle .
\ee
The last line results from the admissible combinations between the quadratic operators imposed
by the action (\ref{Sk}). Only quadratic terms such as $a_{\Bk}\bar a_{\Bk}$, $a_{-\Bk}\bar a_{-\Bk}$,
$a_{\Bk}a_{-\Bk}$, and $\bar a_{\Bk}\bar a_{-\Bk}$, lead to non zero averages. Their corresponding
thermal averages are given by
\bb\nn
\langle a_{\Bk}\bar a_{\Bk}\rangle&=\frac{z+\bar g_{\Bk}}{Z_{\Bk}},
\langle a_{-\Bk}\bar a_{-\Bk}\rangle =\frac{z+g_{\Bk}}{Z_{\Bk}},
\\ \label{eq_aver}
\langle a_{\Bk}a_{-\Bk}\rangle& =-2iu\frac{\sin(k_y)}{Z_{\Bk}},
\langle \bar a_{\Bk}\bar a_{-\Bk}\rangle =2iu\frac{\sin(k_x)}{Z_{\Bk}}.
\ee
Therefore the local density, which is independent of the position, is given by
\bb\nn
\rhop&=\langle n_{\Br}\rangle=
\frac{z}{N}{\sum_{\Bk}}^{'}
\frac{2z+g_{\Bk}+\bar g_{\Bk}}{Z_{\Bk}}
\\ \label{eq_z}
&=\frac{2z}{N}{\sum_{\Bk}}^{'}
\frac{1+z-u(\cos(k_x)+\cos(k_y))-u^2\cos(k_x+k_y)}{Z_{\Bk}}
=z\FZ(z),
\ee
where $\FZ(z)$ can be expressed exactly in terms of complete elliptic integrals in the continuum or asymptotic limit, see in reference \cite{Fortin:2021} the Appendix C for precise details. In particular,
the complete expression at finite temperature is given by
\bb\label{eq_Fz_T}
\FZ(z)=\frac{1}{2z}-\frac{1}{z}\frac{1-z^2+u^2(2+u^2)}{(1+z)^2+u^2(2+u^2)}\frac{K_1}{8\pi^2}
+\frac{1}{z}\frac{u(1-u^2)}{(1+z)^2+u^2(2+u^2)}\frac{K_2}{2\pi^2},
\ee
where $K_1$ and $K_2$ are given by
\bb
K_1=\frac{8\pi}{|\al|}\frac{K(q)}{[(r_1^2-1)(r_2^2-1)]^{1/4}},\;q^2=\frac{1}{2}\left (
1-\frac{(r_1r_2-1)}{\sqrt{(r_1^2-1)(r_2^2-1)}}\right ),
\ee
and
\bb\nn
K_2=\frac{8\pi}{|\al|}\frac{1}{[(r_1^2-1)(r_2^2-1)]^{1/4}}\frac{1+R_0}{1-R_0}\left [
K(q)-\Pi\left (-\frac{(1-R_0)^2}{4R_0},q\right )\right ], 
\\
R_0^2=\frac{(r_1^2-1)(r_2^2-1)}{[(r_1+1)(r_2+1)]^2}.
\ee
$K$ and $\Pi$ are the complete elliptic integrals of the first and third kind, respectively.
The quantities $r_1>r_2>1$ are parametrized by factors $\al$ and $\be$, such that
\bb\nn
\al=\frac{2u(1+z-u^2)}{(1+z)^2+u^2(2+u^2)},\;\be=\frac{2u^2z}{(1+z)^2+u^2(2+u^2)},
\\
r_1=\frac{1+\be+\sqrt{\al^2+2\be(1+\be)}}{\al},
\;
r_2=\frac{1+\be-\sqrt{\al^2+2\be(1+\be)}}{\al}.
\ee
%
\begin{figure*}[!ht]
\centering
\includegraphics[width=0.6\linewidth,clip]{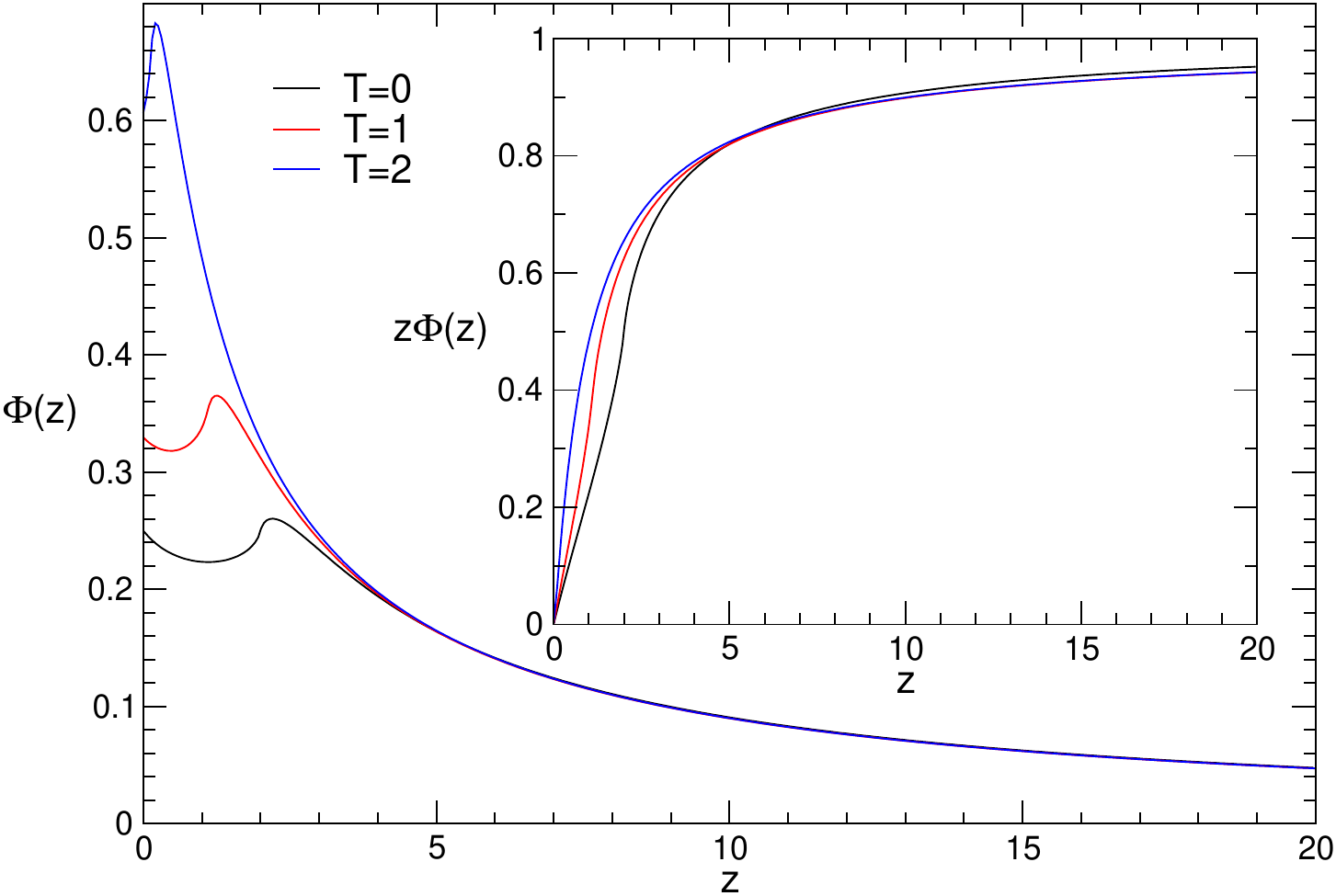}
\caption{
Functions $\FZ(z)$ and $z\FZ(z)$ (inset) for different temperatures ($L=400$). 
$z\FZ(z)$ is positive and monotonic and bounded by unity.
}\label{fig_phi_z}
\end{figure*}
%
In figure \ref{fig_phi_z} is plotted the function $\FZ(z)$ for different temperatures, as
well as $z\FZ(z)$. The latter is bounded by unity for large $z$, or when the concentration of particles
is close to $\rho=1$.
%
\section{Critical properties\label{sect_b}}
%
In presence of dilution, the usual second order critical line for the spin sector has a reduced critical temperature
$T_c(\rhop)$.
%
\begin{figure}[!hb]
\centering
\includegraphics[angle=0,scale=0.7,clip]{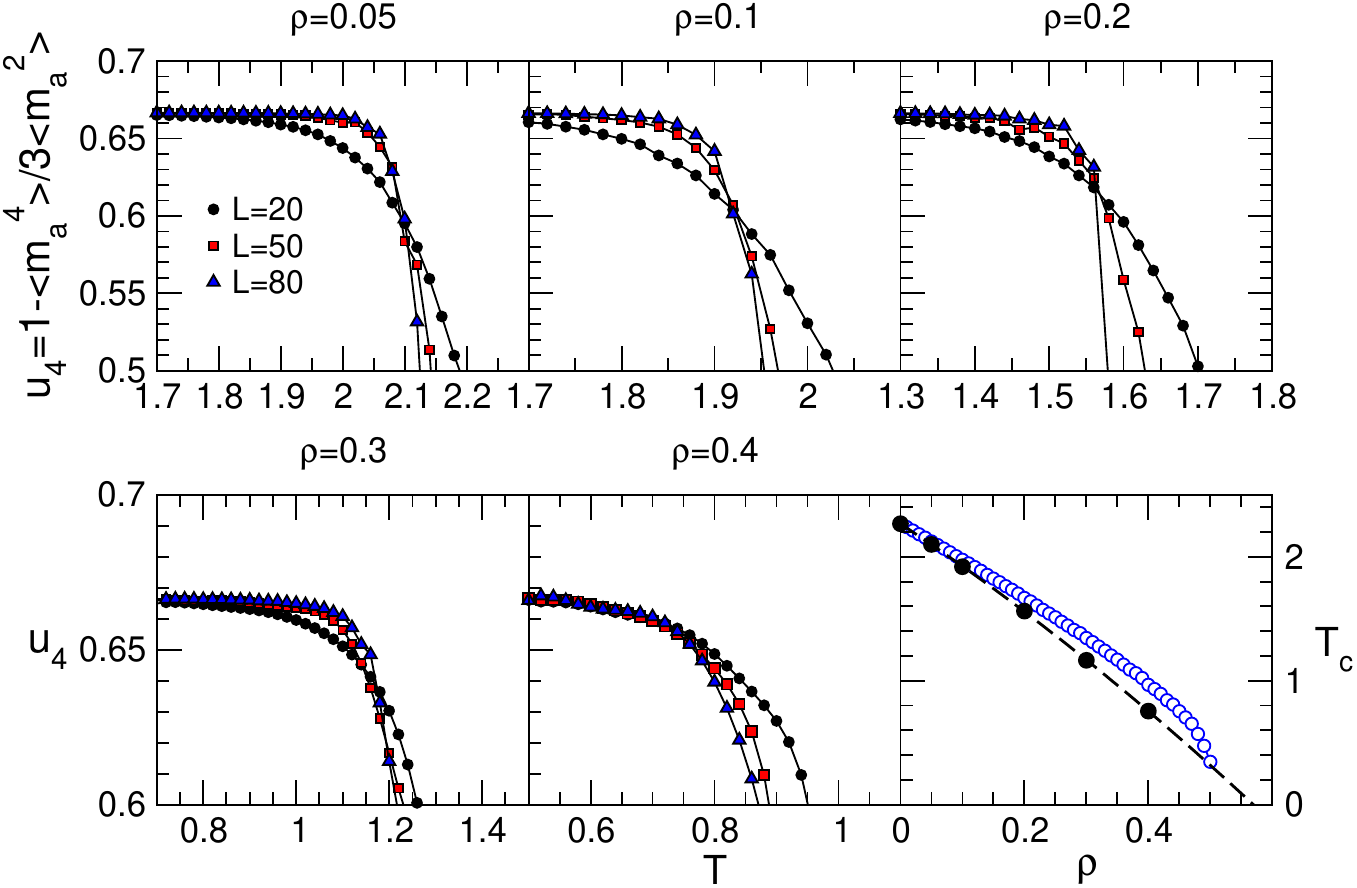}
\caption{
Fourth order cumulant $u_4$ of the magnetization, for different values of $\rhop$ and sizes ($L=20,50,80$).
The value of the critical temperature (blue circles, lower right panel) from the analytical result is compared with the MC simulations (black filled circles) of the finite size estimation of $u_4$, see \eref{eq_u4}. The asymptotic critical temperature for a given concentration is located where the curves for different $L$ intersect.
}\label{fig_MC_u4}
\end{figure}
%
The critical temperature corresponds to the singular value for which $Z_{\Bk}$ in \eref{eq_Zk}
vanishes at long momenta, or when $\Bk=0$. This gives an equation between $u$ and $z$
at $T_c(\rhop)$. In this limit, we can show that 
\bb\label{eq_uc}
Z_{\Bk\simeq 0}=(1+z-2u-u^2)^2=m^2\ge 0,
\ee
where $m$ is defined as the mass of the theory. The critical point is then simply determined by $u=\sqrt{z+2}-1$, with $z$ defined by equation \eref{eq_z}.
The critical temperature is then uniquely determined. When $\rhop=0$, $z=0$ and we have the usual value $T_c(0)=1/\arctanh(\sqrt{2}-1)\simeq 2.269$. When $\rhop$ increases, this value diminishes as more particles destroy the magnetic order until the percolation of large clusters is rendered impossible. Since $u\le 1$, the density for which $T_c=0$ is given by $z=2$, corresponding
to a system half populated by particles, or $\rhop=1/2$, see equation \eref{eq_Fz_T0} in the next section for the exact expression of $\FZ(2)$ in this limit. In figure 
\ref{fig_MC_u4} is plotted the analytical solution $T_c(\rhop)$ as function of $\rhop$, in the last lower right panel with blue circle symbols. For comparison, we have performed MC simulations for the model A and evaluated the fourth order cumulant \cite{Binder:1981} of the magnetization $m_a=\sum_{\Br}\sigma_{\Br}(1-n_{\Br})/\NS$
\bb\label{eq_u4}
u_4=1-\frac{\langle m_a^4\rangle}{3\langle m_a^2\rangle^2},
\ee
for different sizes $L=20,50,80$ as function of the temperature, see figure \ref{fig_MC_u4}. 
For each case, we have performed $10^6$ MC steps, each step corresponding to $N=L^2$ iterations,
starting with a random configuration of particles.
The asymptotic value $T_c(\rhop)$ when $N\rightarrow\infty$ is localized where all the curves intersect, and this value is
displayed as the black filled circles on the last lower right panel of the figure, for each concentration $\rhop$.
Although the analytical results predicts that $T_c$ vanishes at $\rhop=1/2$, the MC results suggests
that the critical temperature vanishes at $\rhop\simeq 0.570$, after a quadratic fit of the data was
performed, see the black dashed line. A linear fit gives instead $\rhop\simeq 0.606$. It is however difficult to extract a precise value of $T_c$ when the density of particles is close to $1/2$ and the analytical result shows important non linearity deviation near this value. To put these results into
perspective, we can mention that the percolation threshold for particles on a square lattice is given with high accuracy by $\rhop\simeq 0.593$ \cite{Ziman:1979,Newman:2000}. For mobile electrons on a square lattice with Ising spins, using the quadratic approximation of the Grassmann action and self-consistent equations for the fermionic Green functions, we found that the critical density is instead $\rhop=2/3\simeq 0.667$ \cite{Fortin:2021}, which is a larger value.
%
%
\section{Entropy and specific heat\label{sect_c}}
%
The free energy density per site in the Canonical Ensemble is given by the expression
\bb\label{eq_F}
\frac{F}{N}=-k_BT(1-\rhop)\Big [\log 2+2\log \cosh \beta J\Big ]
+k_BT\rhop\log z
-\frac{k_BT}{N}{\sum_{\Bk}}^{'}
\log Z_{\Bk}.
\ee
The entropy per site is defined by $S(T)=-N^{-1}\partial F/\partial T$. At high temperature,
the previous expression can be simplified ($u\simeq 0$) by expanding the free energy and keeping
the leading terms
\bb
S(T\gg 1)/k_B\simeq (1-\rhop)\log 2+\rhop\log z+\log (1+z).
\ee
The value of $z$ in this limit, given by the implicit equation \eref{eq_z}, yields the simple relation
$z=\rhop/(1-\rhop)$. Therefore, we deduce that the entropy is simply given by the sum
of the spin contribution, $\log 2$, on each site where a spin is present, and the entropy for $\NP$ free particles 
\bb\label{eq_S_Tlarge}
S(T\gg 1)/k_B\simeq (1-\rhop)\log 2-\rhop\log \rhop-(1-\rhop)\log (1-\rhop).
\ee
The particle contribution corresponds to the number of ways to put randomly $\NP$ particles on a
square lattice with each site containing at most one particle. This contribution is equivalent to the logarithm of the combinatorial factor ${N\choose\NP}$ in the thermodynamical limit $N\rightarrow\infty$
using the Stirling formula.
%
\begin{figure*}[!ht]
\centering
\begin{subfigure}[b]{0.45\linewidth}
\caption{}
\includegraphics[width=\linewidth,clip]{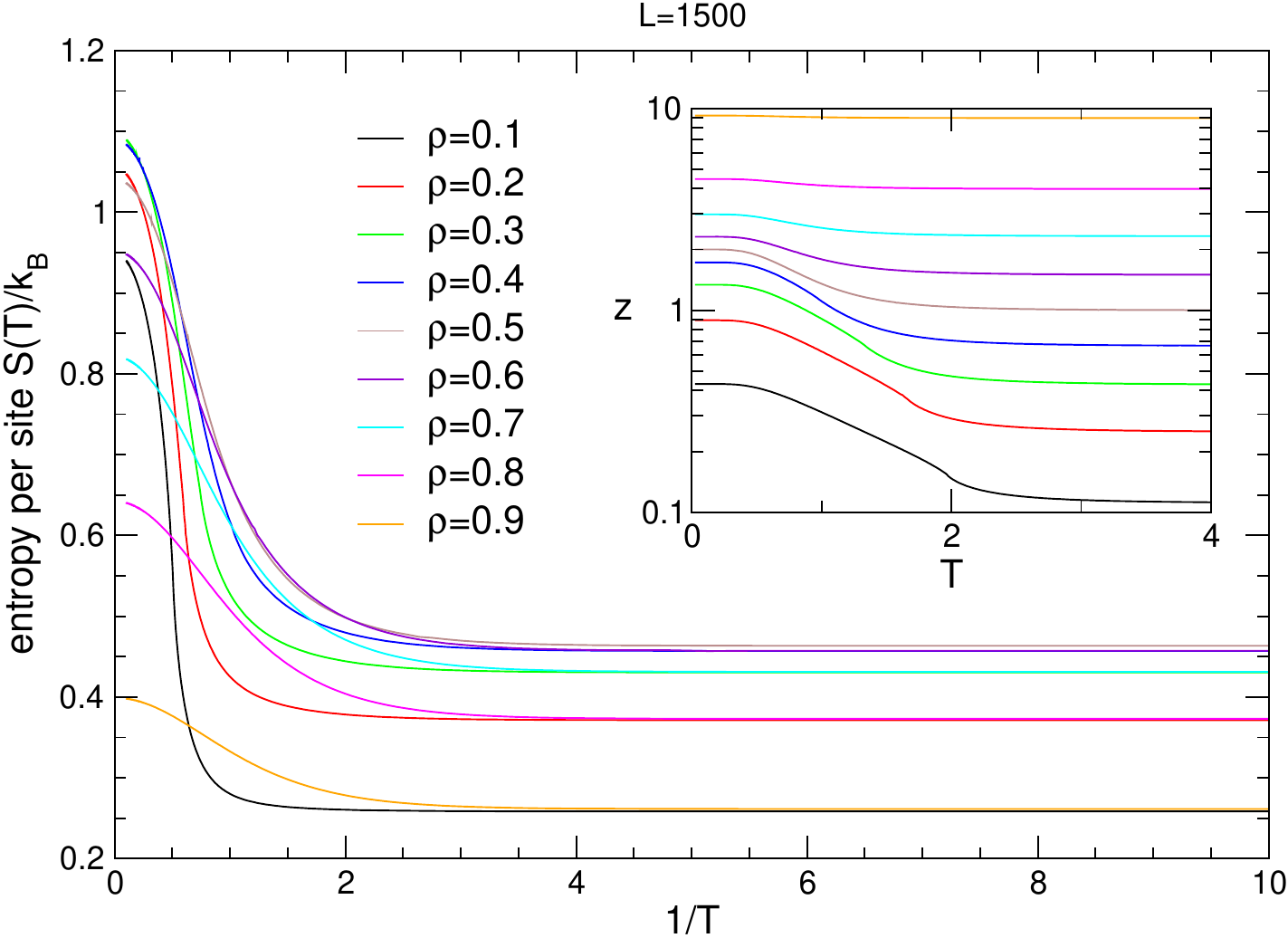}
\end{subfigure}
\begin{subfigure}[b]{0.45\linewidth}
\caption{}
\includegraphics[width=\linewidth,clip]{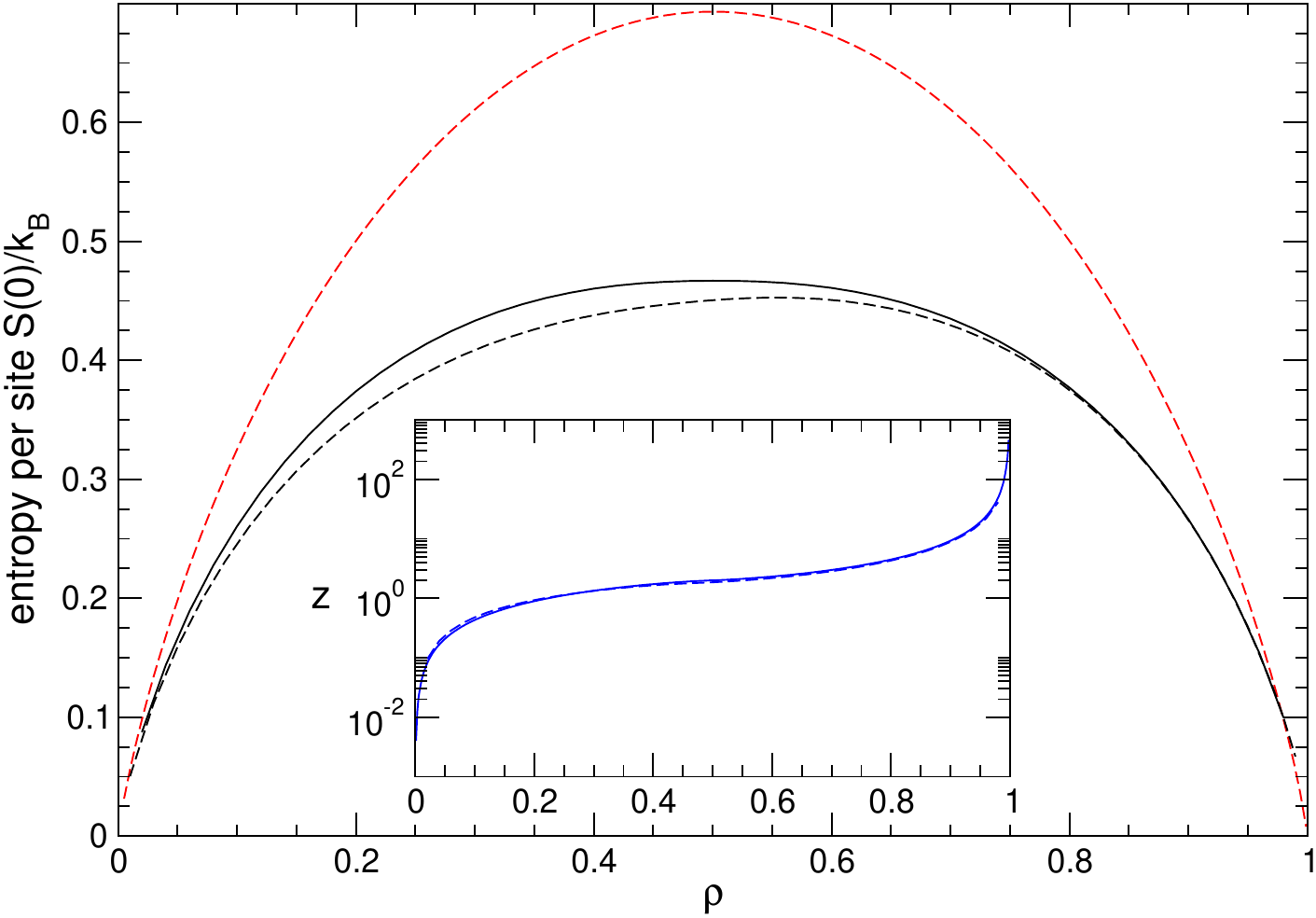}
\end{subfigure}
\caption{
(a) Entropy per site as function of the temperature $T$ for different values of $\rho$ ($L=1500$), with
the variation of the activity $z$ with $T$ in inset.
(b) Entropy per site at zero temperature as function of the density of particles $\rhop$. 
The black dashed line represents the entropy with the quartic corrections, see \ref{app_1}. 
For comparison, the red dashed line represents the free entropy $S/k_B=-\rhop\log \rhop-(1-\rhop)\log (1-\rhop)$ if the particles were all independent. In inset is the corresponding value of $z$.
}\label{fig_entropy}
\end{figure*}
%
In figure \ref{fig_entropy}(a) is represented the temperature dependent entropy from expression
\eref{eq_F}, and for different concentrations of particles. At high temperature, the entropy
reaches the value given by \eref{eq_S_Tlarge}, and a crossover around $T\simeq 1$ is observed below
which the entropy tends to a low temperature macroscopic residual value.

\begin{figure*}[!ht]
\centering
\begin{subfigure}[b]{0.45\linewidth}
\includegraphics[width=0.65\linewidth,clip]{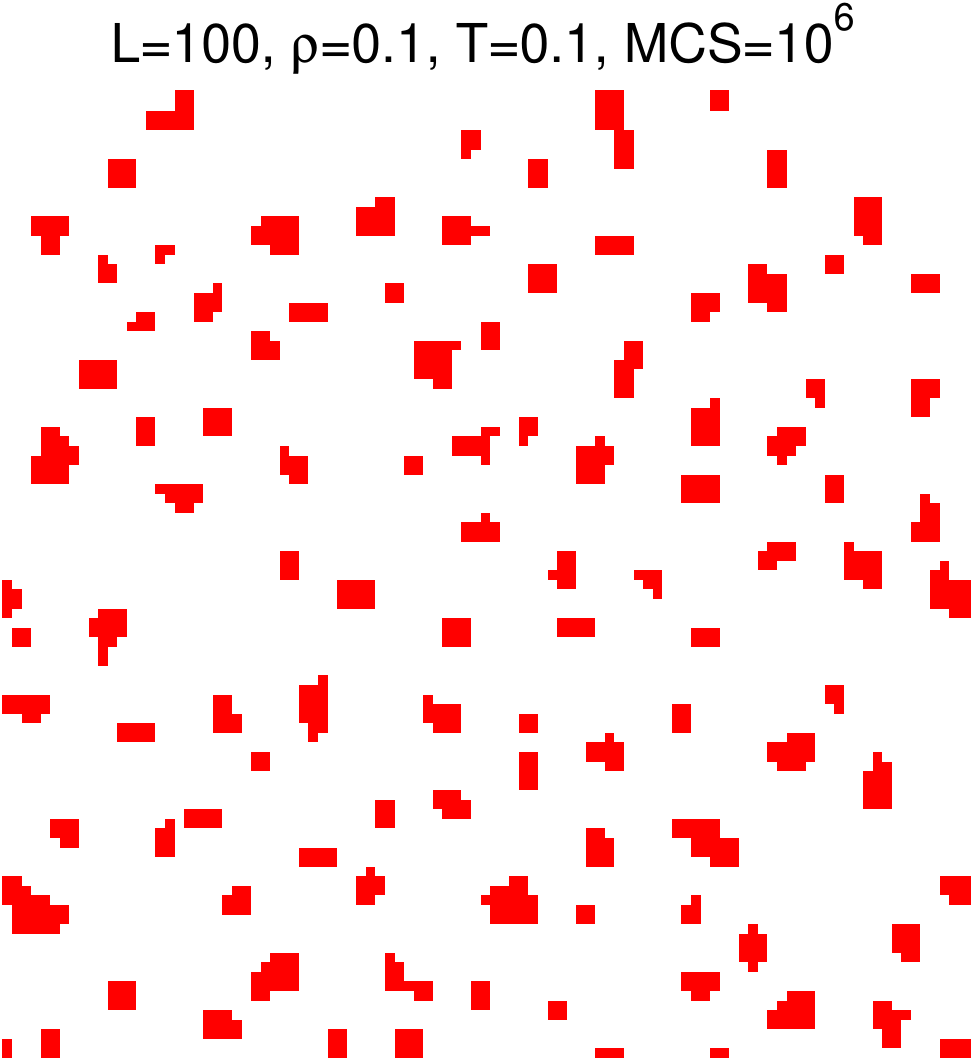}
\end{subfigure}
\begin{subfigure}[b]{0.45\linewidth}
\includegraphics[width=0.65\linewidth,clip]{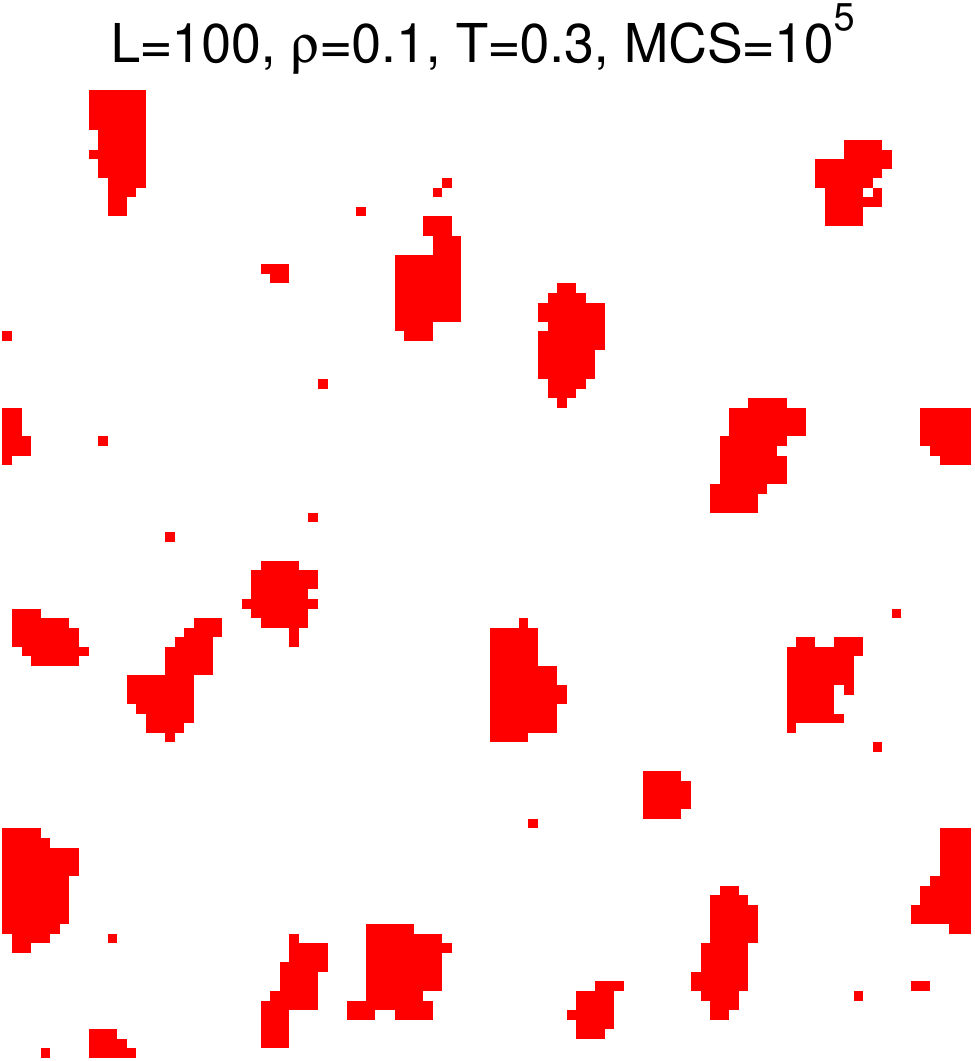}
\end{subfigure}
\begin{subfigure}[b]{0.45\linewidth}
\includegraphics[width=0.65\linewidth,clip]{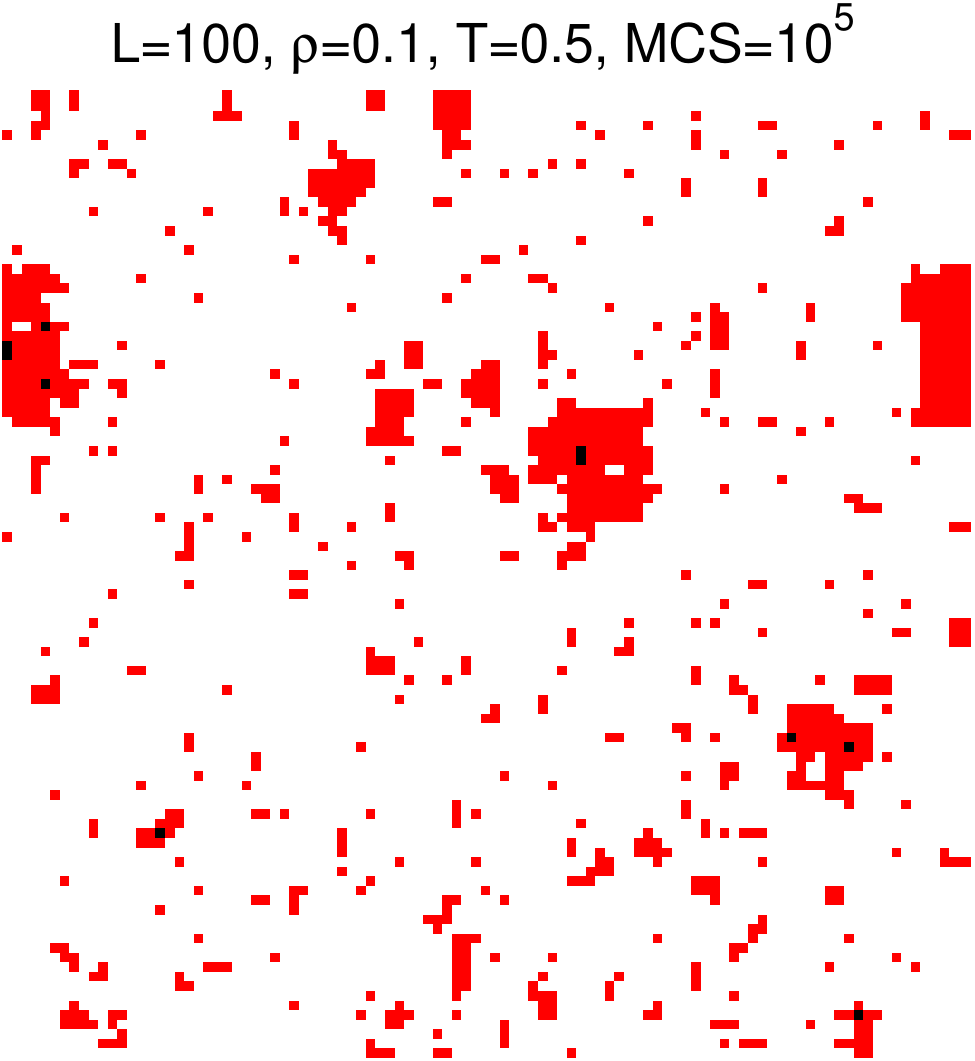}
\end{subfigure}
\begin{subfigure}[b]{0.45\linewidth}
\includegraphics[width=0.65\linewidth,clip]{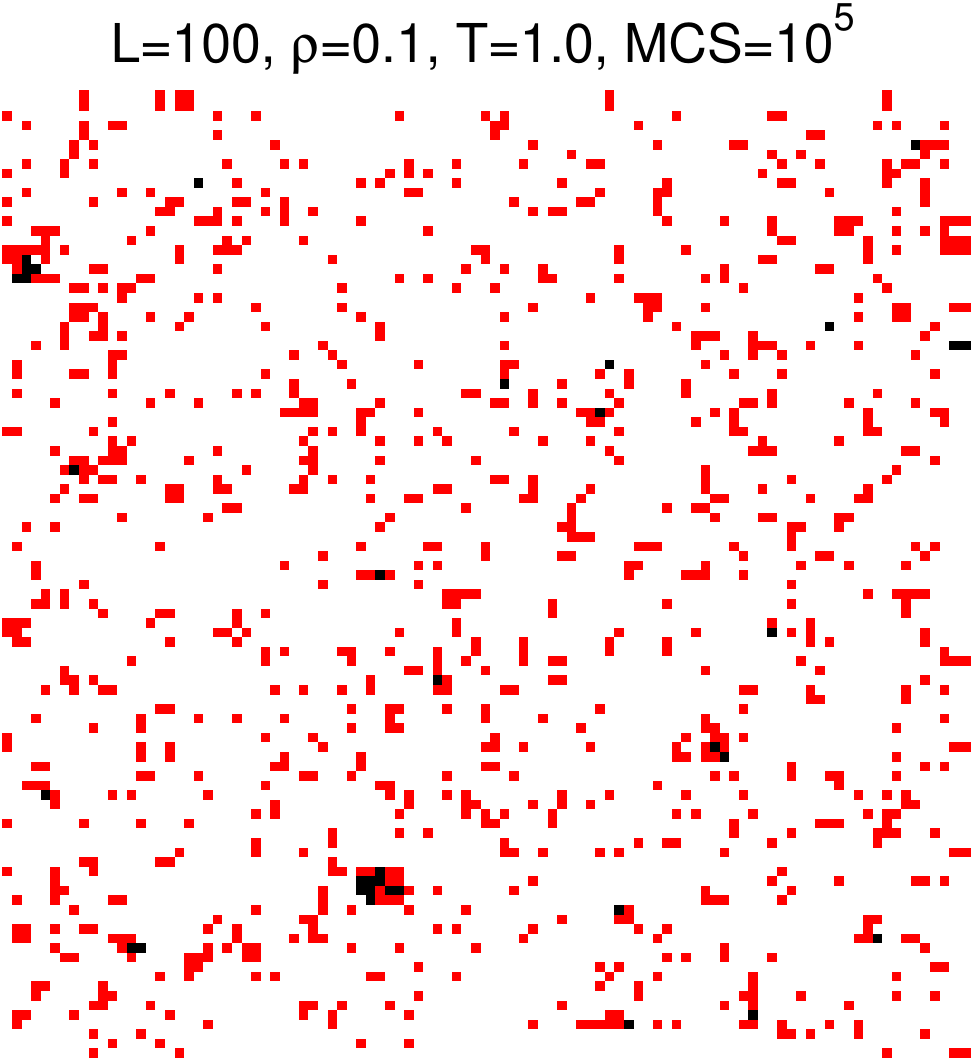}
\end{subfigure}
\begin{subfigure}[b]{0.45\linewidth}
\includegraphics[width=0.65\linewidth,clip]{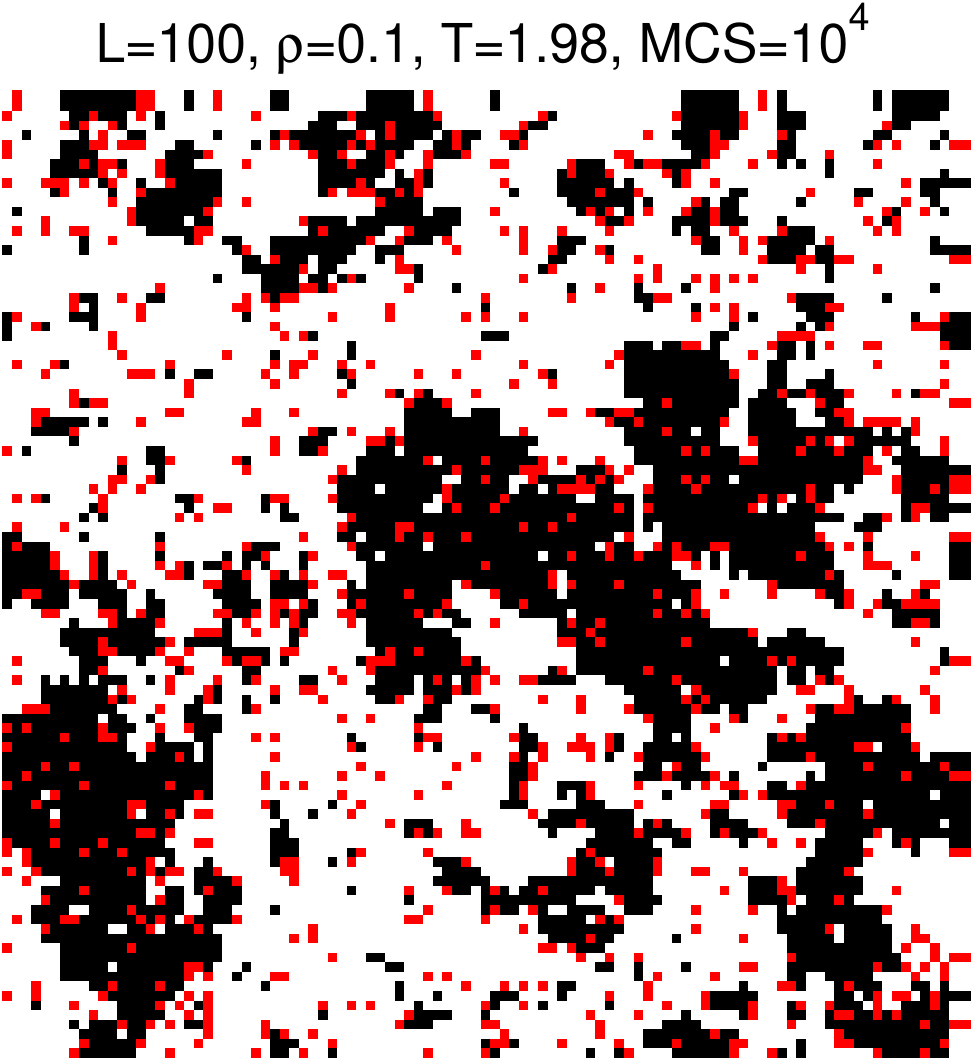}
\end{subfigure}
\begin{subfigure}[b]{0.45\linewidth}
\includegraphics[width=0.65\linewidth,clip]{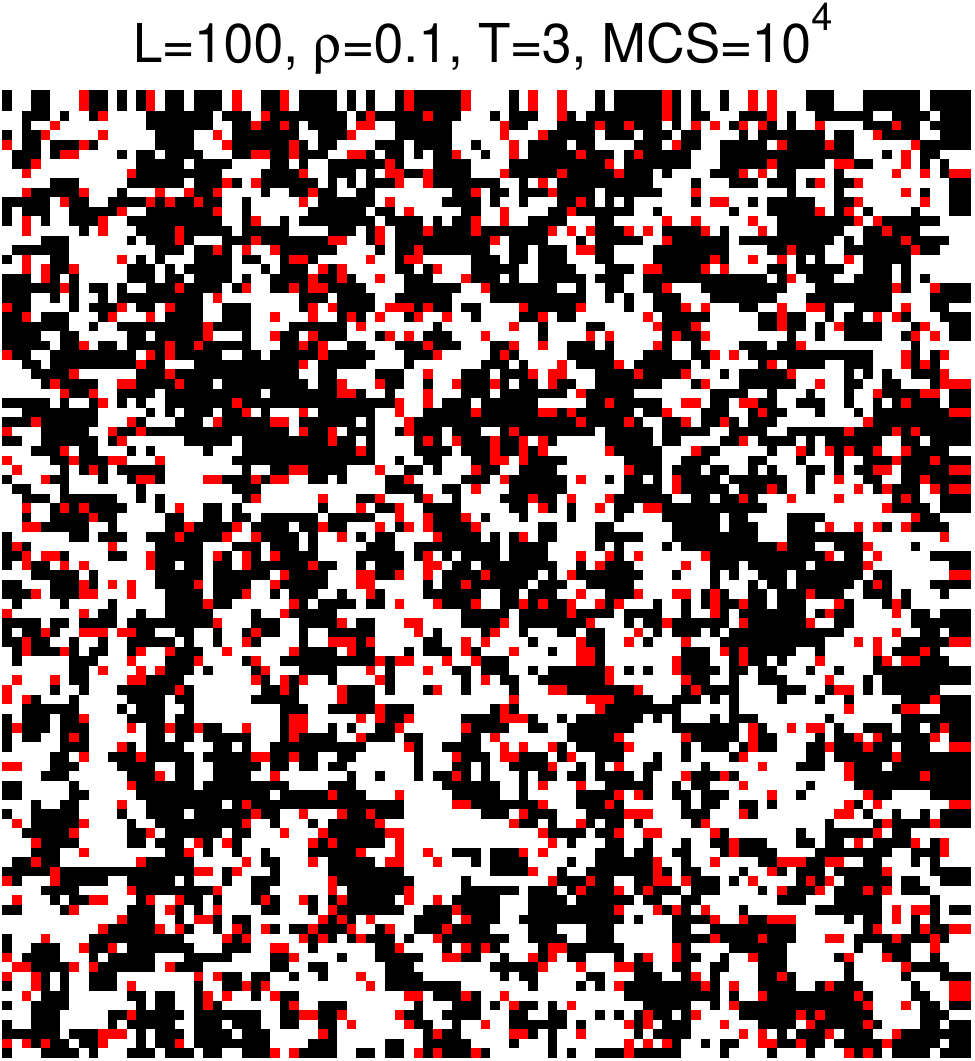}
\end{subfigure}
\caption{
Spin and particle MC configurations for different temperatures, model A: the white squares represent spins up, black squares spins downs, and red squares the particle positions. For each temperature,
the initial particle and spin configuration of the MC simulation is chosen randomly.
}\label{fig_config_MC}
\end{figure*}

\begin{figure*}[!ht]
\centering
\begin{subfigure}[b]{0.45\linewidth}
\includegraphics[width=0.75\linewidth,clip]{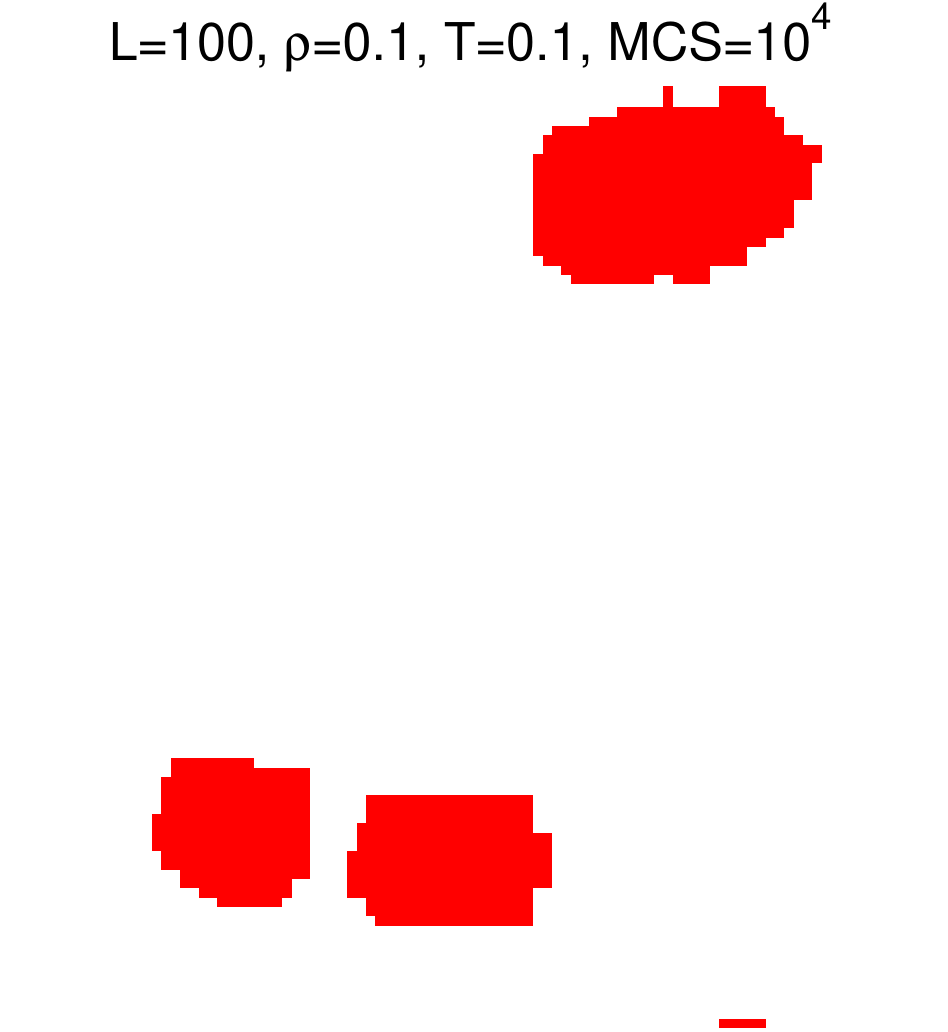}
\end{subfigure}
\begin{subfigure}[b]{0.45\linewidth}
\includegraphics[width=0.75\linewidth,clip]{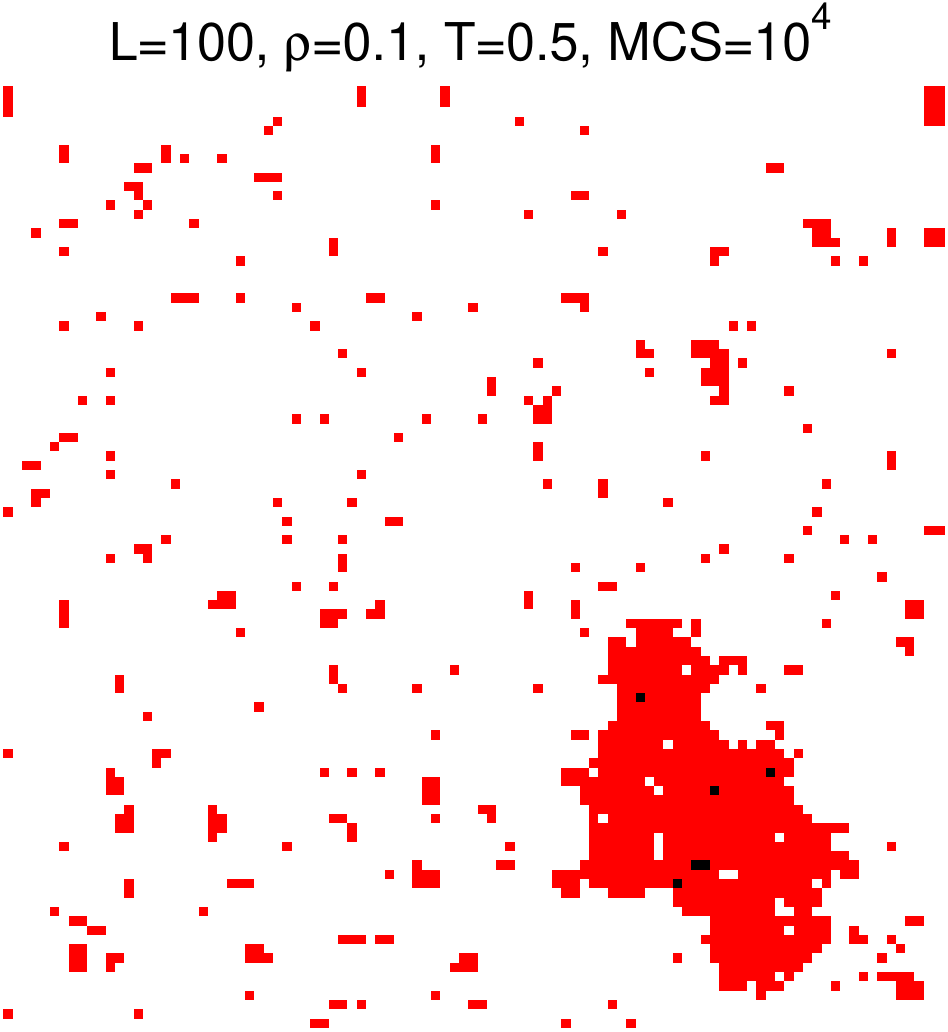}
\end{subfigure}
\caption{
Spin and particle MC configurations for different temperatures, model B. The initial particle and spin configuration of the MC simulation is chosen randomly.
}\label{fig_config_MC_B}
\end{figure*}

In the low temperature regime, this residual entropy is less obvious to determine analytically. In absence of any particle, the entropy is simply zero since there are only two possible ground states with all spins up
or all spins down. However, in presence of particles, this entropy is not zero as the particles tend naively to arrange themselves in clusters with non-trivial geometry in order to minimize the energy and maximize the coupling between spins, either with the models A or B.
After expanding equation \eref{eq_F} at small $T$ and keeping the dominant terms which are not exponentially small, we find
\bb\nn
S(0)/k_B&=-(1-\rhop)\log 2-\rhop\log z
\\ \label{eq_S0}
&+\frac{1}{N}{\sum_{\Bk}}{'}
\log \left (
z^2+2z+4-2z\left [ \cos(k_x)+\cos(k_y)+\cos(k_x+k_y) \right ] \right ),
\ee
where $z$ is obtained from \eref{eq_z} at zero temperature. We can notice that if $\rhop=0$,
then $z=0$, and $\rhop\log z\rightarrow 0$ since $\FZ(z)$ is finite near $z=0$, see figure \ref{fig_phi_z}. Then the expression
above reduces to $S(0)=0$ as expected. Function $\FZ(z)$
has the following expression at zero temperature in the thermodynamical limit, after setting
$u=1$ in \eref{eq_Fz_T}
\bb\nn
\FZ(z)=\frac{1}{2z}+\frac{(z-2)}{\pi z\sqrt{z^2-4}}K\left (\sqrt{\frac{1-R}{2}}\right ),
\\
R=\frac{z^4+4z^3+16z+16}{(2+z)^2|z^2-4|}.
\ee
In this formula, $K$ is the complete elliptic integral of the first kind with imaginary modulus
since the quantity $(1-R)/2$ in the square root of the previous expression is found to be negative. We obtain in particular
\bb\nn
\FZ(z\le 2)=\frac{1}{2z}-\frac{1}{\pi z}\sqrt{\frac{2-z}{2+z}}K\left (i
\sqrt{\frac{z^3(4+z)}{(2+z)^3(2-z)}}\right ),
\\ \label{eq_Fz_T0}
\FZ(z\ge 2)=\frac{1}{2z}+\frac{1}{\pi z}\sqrt{\frac{z-2}{z+2}}K\left (i
\sqrt{\frac{16(1+z)}{(2+z)^3(z-2)}}\right ).
\ee
At the special singular point $z=2$, we have simply $\FZ(2)=1/4$. This corresponds graphically to the peak of the function $\FZ(z)$ displayed in figure \ref{fig_phi_z}.
In figure \ref{fig_entropy}(b) is plotted the entropy $S(0)/k_B$ at zero temperature as function of 
the particle density. We have used \eref{eq_Fz_T0} to determine $z$ and computed the entropy from \eref{eq_S0} with $L=2500$ to evaluate the finite sum. The entropy is symmetrical around $\rhop\simeq 0.5$ where it reaches its maximum $S(0)\simeq 0.461\, k_B$. It is given in the thermodynamical limit by the integral
\bb\fl
S(0)/k_B=\frac{1}{8\pi^2}\int_{-\pi}^{\pi}\int_{-\pi}^{\pi}dk_xdk_y
\log\left [3-\cos(k_x)-\cos(k_y)-\cos(k_x+k_y)\right ]
\simeq 0.46108.
\ee
The black dashed curve in figure \ref{fig_entropy}(b) corresponds to the contribution of the quartic
terms in the main action $\SP$ defined in \eref{ZGC}, see \ref{app_1} for analytical details.
The improved value of the entropy is now slightly lower and non symmetrical about $\rhop=0.5$ but close to the approximated entropy with quadratic terms only. In addition, the residual entropy is lower than the one for $\NP$ hard core independent particles, see red dashed line on the same figure. This nonzero
entropy at first could be attributed to a crystallization effect at low temperature, where particles tend to form independent clusters or one giant unique cluster, in models A and B, as we can see in figures \ref{fig_config_MC} and \ref{fig_config_MC_B} respectively.
%
%
%
%
From a random configuration of particles, the unique cluster is rapidly formed in the model
B, as expected from nonlocal hopping particles, see for example the configuration at $T=0.1$ in figure \ref{fig_config_MC_B}. In the model A, starting from a random configuration of particles in the low temperature regime, small clusters are formed principally since such configurations are metastable states. At zero temperature, these configurations are stable, unless a cooling procedure is performed from a high temperature. At slightly higher temperatures, for example $T=0.3$ and $T=0.5$, the mobility of the particles, or diffusivity, is larger and bigger clusters can be formed after some time. Close to the critical point $T\simeq 1.98$ for $\rhop=0.1$, or even at high temperature regimes, $T=3$, many particles are located near the domain walls. 

In order to probe the existence of phases, in the spin and impurity sectors, we evaluate numerically the specific heat by considering the energy fluctuations
%
\begin{figure*}[!ht]
\centering
\begin{subfigure}[b]{0.48\linewidth}
\caption{}
\includegraphics[width=\linewidth,clip]{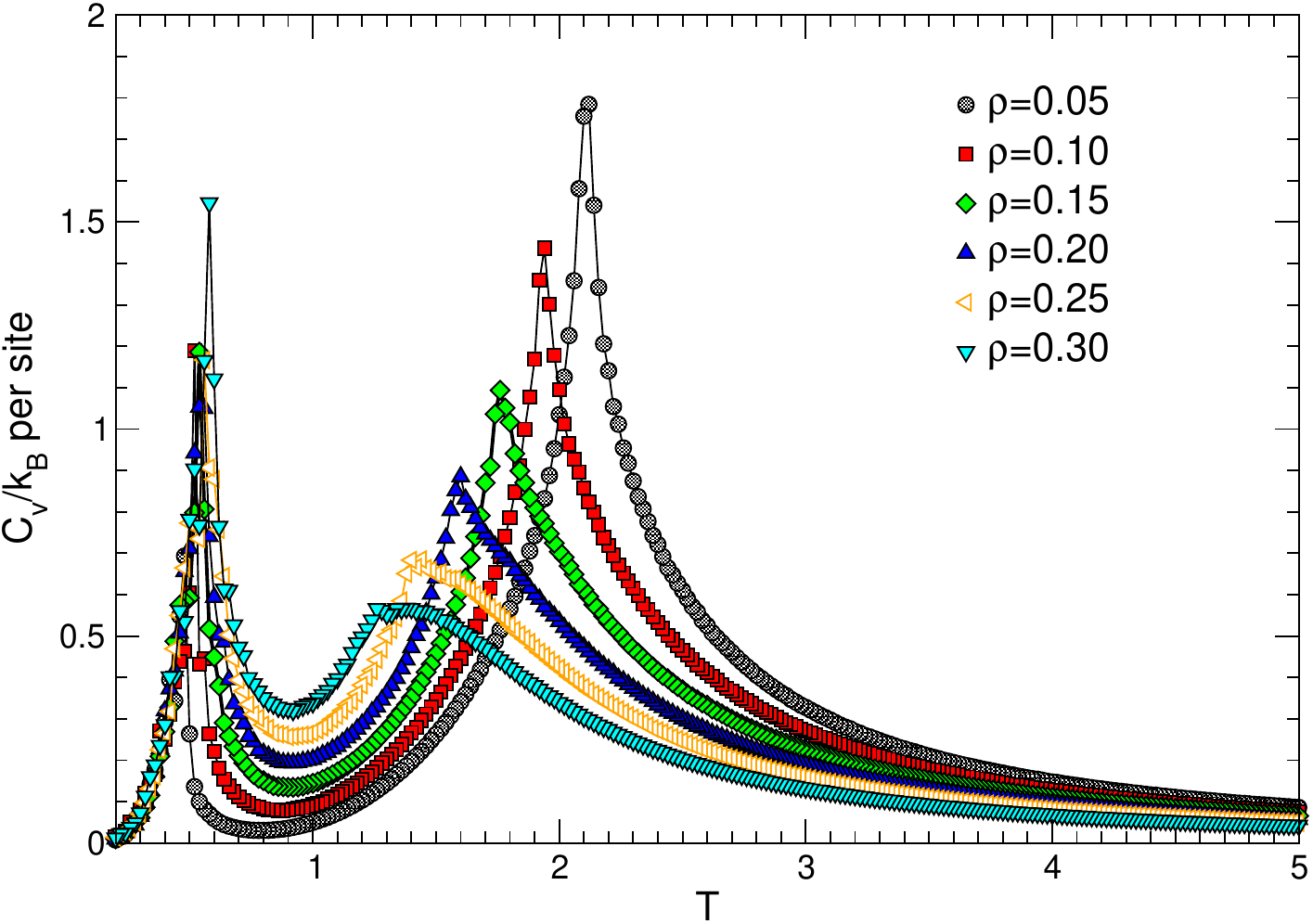}
\end{subfigure}
\begin{subfigure}[b]{0.48\linewidth}
\caption{}
\includegraphics[width=\linewidth,clip]{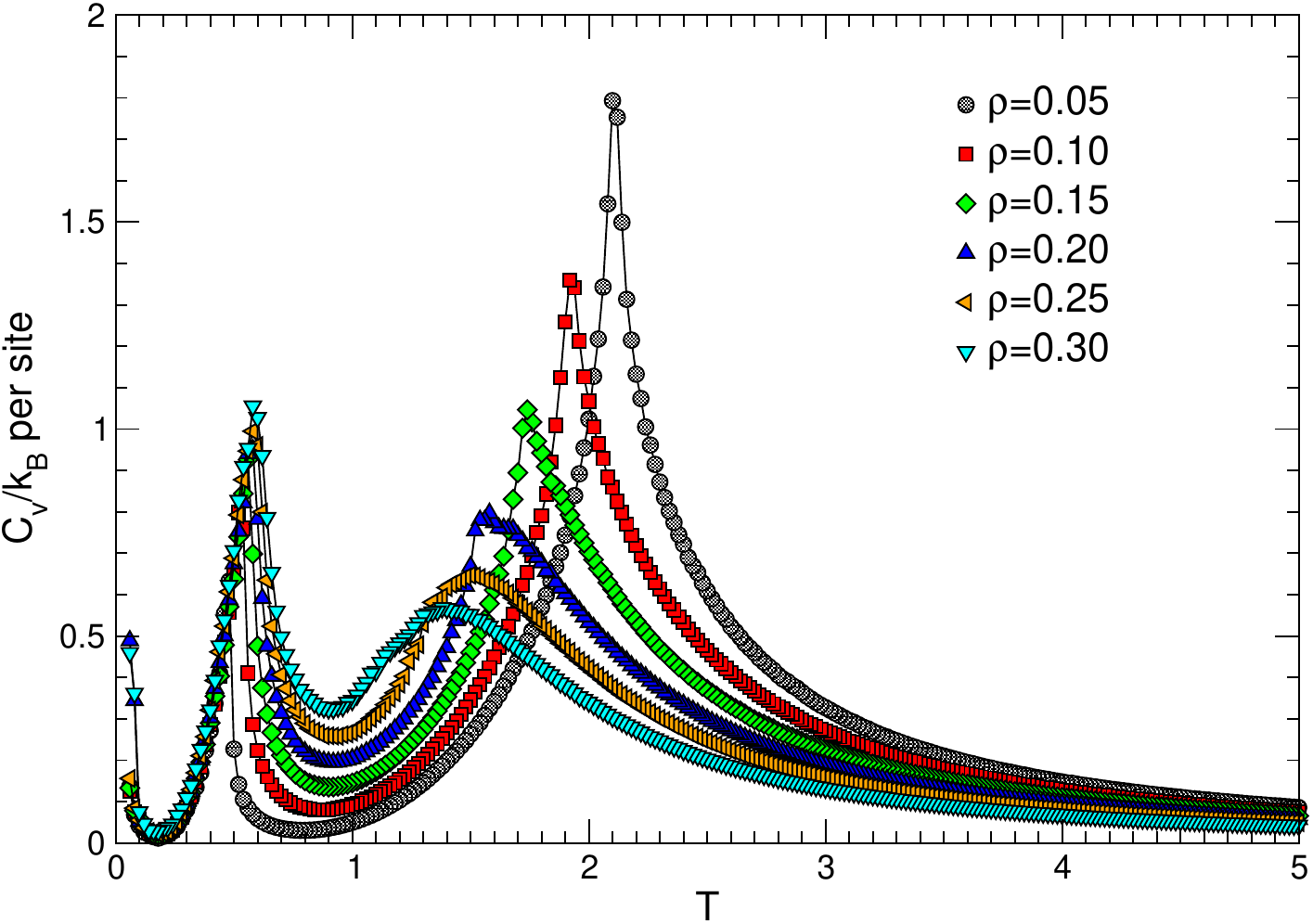}
\end{subfigure}
\caption{
(a) Specific heat per site for $L=100$ with $10^6$ MCS, model A, as function of the temperature.
(b) Specific heat per site, model B, as function of the temperature. In both cases the numerical data were recorded by decreasing the temperature (cooling).
}\label{fig_cv_MC}
\end{figure*}
%
%
\begin{figure*}[!ht]
\centering
\includegraphics[width=0.6\linewidth,clip]{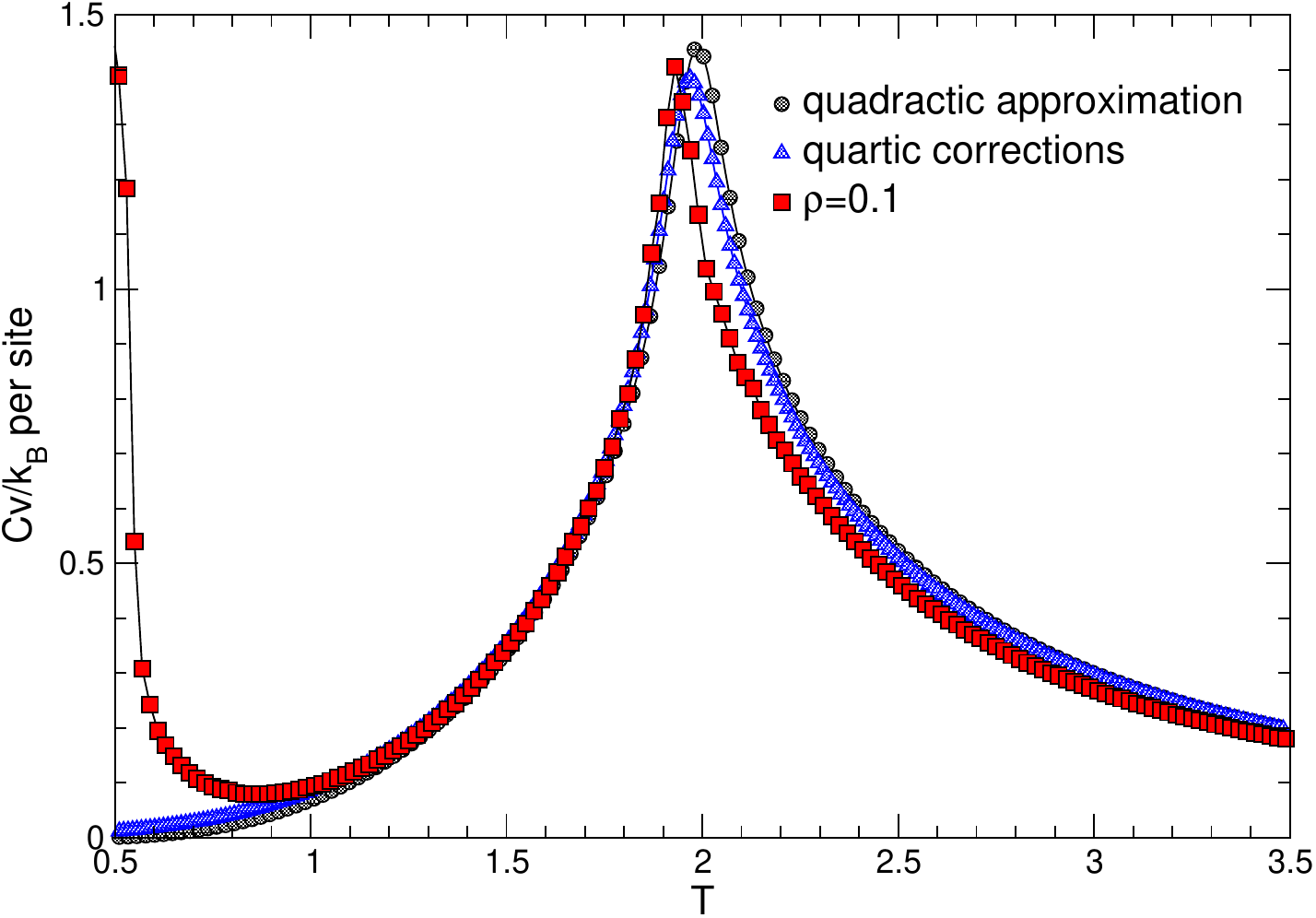}
\caption{
Comparison of $C_v$ for $\rhop=0.1$ with the equations \eref{eq_F} and \eref{eq_cv}, using quadratic and quartic approximations from \ref{app_1}, model A.
}\label{fig_cv_MC_fit}
\end{figure*}
%
%
\bb\label{eq_cv}
C_v(T)=k_B\beta^2\left (\langle \HS^2\rangle-\langle \HS\rangle^2\right )
\ee
or $C_v=-T\partial^2F/\partial T^2$. From the expression \eref{eq_cv} we can in principle deduce the 
entropy at finite temperature from the exact value at infinite temperature 
\bb\label{eq_ST}
S(T)=S(\infty)-\int_T^{\infty}\frac{C_v(T')}{T'}dT',
\ee
with $S(\infty)$ given by \eref{eq_S_Tlarge}.
In figures \ref{fig_cv_MC}(a) and (b) we have plotted the specific heat from the MC results, for several particle concentrations, in a system of size $L=100$ with $10^6$ MCS, and for the two models A and B respectively. We have evaluated $C_v$ up to $T=10$ for fitting analytically $C_v$ at large temperature in order to compute the entropy, but we have restricted the figures at a maximum temperature $T=5$ for display purpose. In figure \ref{fig_cv_MC_fit}, for $\rhop=0.1$, is plotted the analytical result from \eref{eq_F}, as well as the corrective contributions from the quartic term in the action $\SP$. The corrective terms agree well with the numerical results around the critical point for this concentration, however $C_v$ is slightly
larger than the MC data at higher temperature. 

In the low temperature regime, $T\lesssim 1$, the 
MC data in figures \ref{fig_cv_MC}(a) and (b) present a peak structure at low temperature, around $T\simeq 0.5$ for all defect concentrations,
which we attribute to the formation of particle clusters when they crystallize. These clusters are usually not extended, see figure \ref{fig_config_MC_B} for example, and we may assume that this peak 
represent an anomaly but they possess a sharp structure which is more compatible to
the presence of large energy fluctuations associated with the formation of metastable states. 
The amplitude of this peak is slightly dependent of the model A or B chosen: In model A the peak amplitude is larger than in model B which displays additional fluctuations at temperatures $T<0.2$
due to a possible residual particle activity.
%
\begin{figure*}[!ht]
\centering
\includegraphics[width=0.6\linewidth,clip]{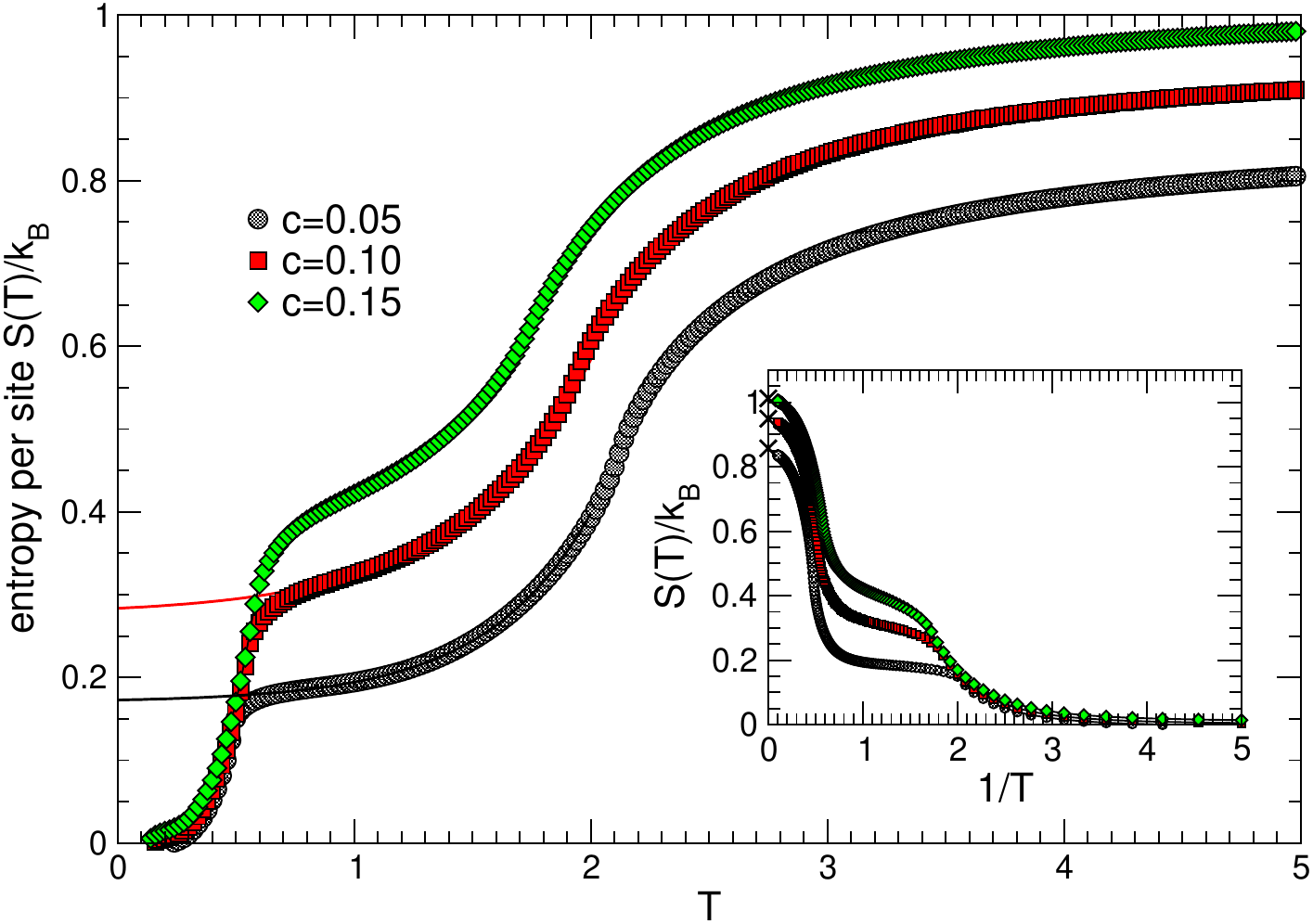}
\caption{Low temperature entropy deduced from the MC data of figure \ref{fig_cv_MC}(b) down to $T=0.1$ for $\rhop=0.05,0.1$ and $0.2$, using the integration \eref{eq_ST} and a fit
at high temperature of the specific heat. The cross symbols in the inset are deduced from the exact expression (\ref{eq_S_Tlarge}) at infinite temperature.
}\label{fig_MC_ST}
\end{figure*}
%
These peaks are however not accounted for by the theory which does not present such a peak in the analytical expressions.
However we can numerically compute the contribution of this peak to the entropy (model B) as function of $T$ from the data of
figure \ref{fig_cv_MC}(b) using equation \eref{eq_ST}. Since we know the exact expression of
the entropy at infinite temperature (\ref{eq_S_Tlarge}), an analytical fit of the form $C_v\simeq
T^{-b}\e^{\frac{a}{T}}$ is performed from $T\gtrsim 2$ to
$T=10$. The resulting expression is integrated from $T=5$ to $T=\infty$, then we use
the data between $T=5$ and a reduced temperature $T$ to extract the entropy by integration, see figure \ref{fig_MC_ST}.
We observe that the entropy tends to vanish at $T\simeq 0.1$ for the three concentrations considered, within approximation errors since the entropy becomes negative below this limit. We conclude that the low temperature peak that we observe around $T=0.5$ is compatible with the existence of a unique ground state or giant cluster. In figure \ref{fig_MC_ST}, we also have extrapolated the entropy for two concentration values (solid lines black and red) and measured the zero temperature entropy without the effect of the first peak at
$T\simeq 0.5$. For $\rhop=0.05$, we find that $S(0)/k_B\simeq 0.173$, close to the theoretical value $0.160$, and for $\rhop=0.1$, we find that $S(0)/k_B\simeq 0.283$, close to $0.260$. This shows that
the first peak contribution corresponds almost exactly to the excess of entropy that was predicted by the theory.
%
\section{Two-points correlation functions\label{sect_d}}
%
We would like to evaluate the connected two-points correlation function $\CC(\Br,\Br')=\langle n_{\Br}n_{\Br'}\rangle-\langle n_{\Br}\rangle\langle n_{\Br'}\rangle$ between the particle densities at two points $\Br\ne\Br'$. We consider as before the Fourier transformation to evaluate the pair correlations with the 
action (\ref{Sk})
\bb\nn
\CC(\Br,\Br')
=\frac{z^2}{N^2}{\sum_{\Bk,\Bk',\Bk'',\Bk'''}}^{'}
\Big \langle
\Big (a_{\Bk} \e^{i\Bk.\Br}+a_{-\Bk} \e^{-i\Bk.\Br}\Big )
\left (\bar a_{\Bk'} \e^{-i\Bk'.\Br}+\bar a_{-\Bk'} \e^{i\Bk'.\Br}\right )
\\
\times\left (a_{\Bk''} \e^{i\Bk''.\Br'}+a_{-\Bk''} \e^{-i\Bk''.\Br'}\right )
\left (\bar a_{\Bk'''} \e^{-i\Bk'''.\Br'}+\bar a_{-\Bk'''} \e^{i\Bk'''.\Br'}\right )
\Big \rangle-\rhop^2.
\ee
There are three possible contractions between the four different pairs, with one giving simply
the disconnected term $\rhop^2$, independent of the position.
We find that the dominant contributions, up to $1/N$ corrections, yield
\bb\label{eq_G}
\CC(\Br,\Br')=-\frac{z^2}{N^2}{\sum_{\Bk,\Bk'}}
{\vphantom{\sum}}'
\Big [
2\sin(\Bk.[\Br-\Br'])
\langle 
a_{\Bk}a_{-\Bk}\rangle .
2\sin(\Bk'.[\Br-\Br'])
\langle\bar a_{\Bk'}\bar a_{-\Bk'}\rangle
\\ \nn
+\Big \langle a_{\Bk}\bar a_{\Bk} \e^{i\Bk.(\Br-\Br')}+a_{-\Bk}\bar a_{-\Bk} \e^{-i\Bk.(\Br-\Br')}
\Big \rangle
\Big \langle a_{\Bk'}\bar a_{\Bk'} \e^{-i\Bk'.(\Br-\Br')}+a_{-\Bk'}\bar a_{-\Bk'} \e^{i\Bk'.(\Br-\Br')}
\Big \rangle
\Big ].
\ee
The correlation function depends therefore only on the distance $\Bd=\Br'-\Br$ between the two points, or 
$\CC(\Br,\Br')=\CC(\Bd)=\CC(d_x,d_y)$. Using the rules (\ref{eq_aver}), we obtain, after integrating over the Grassmann variables
\bb\label{eq_Gd}
\CC(\Bd)=-\frac{16z^2u^2}{N^2}
{\sum_{\Bk}}^{'}
\frac{\sin(k_x)\sin(\Bk.\Bd)}{Z_{\Bk}}
{\sum_{\Bk'}}^{'}
\frac{\sin(k'_y)\sin(\Bk'.\Bd)}{Z_{\Bk'}}
\\ \nn
-\frac{z^2}{N^2}
{\sum_{\Bk}}^{'}
\Big (
\frac{z+g_{\Bk}}{Z_{\Bk}}\e^{i\Bk.\Bd}
+\frac{z+\bar g_{\Bk}}{Z_{\Bk}}\e^{-i\Bk.\Bd} \Big )
{\sum_{\Bk'}}^{'}
\Big (
\frac{z+g_{\Bk'}}{Z_{\Bk'}}\e^{-i\Bk'.\Bd}
+\frac{z+\bar g_{\Bk'}}{Z_{\Bk'}}\e^{i\Bk'.\Bd} \Big ).
\ee
This function is symmetric by inversion $\CC(-\Bd)=\CC(\Bd)$. The expression is also symmetric
along the diagonal axis, $\CC(d_x,d_y)=\CC(d_y,d_x)$. 
We can also define a susceptibility $\chi={\sum_{\Bd\ne 0}}\CC(\Bd)$ from the previous expression using the relation $\sum_{\Bd\ne 0}\e^{i\Bk.\Bd}=N\delta_{\Bk,\bO}-1$, and we find after some algebra the leading terms
\bb\label{eq_chi}
\chi=\rho^2
-\frac{8z^2u^2}{N}{\sum_{\Bk}}^{'}\frac{\sin(k_x)\sin(k_y)}{Z_{\Bk}^2}
-\frac{z^2}{N}{\sum_{\Bk}}^{'}\frac{(z+g_{\Bk})^2+(z+\bar g_{\Bk})^2}{Z_{\Bk}^2}.
\ee
%
%
\begin{figure*}[!ht]
\centering
\begin{subfigure}[b]{0.45\linewidth}
\includegraphics[width=0.95\linewidth,clip]{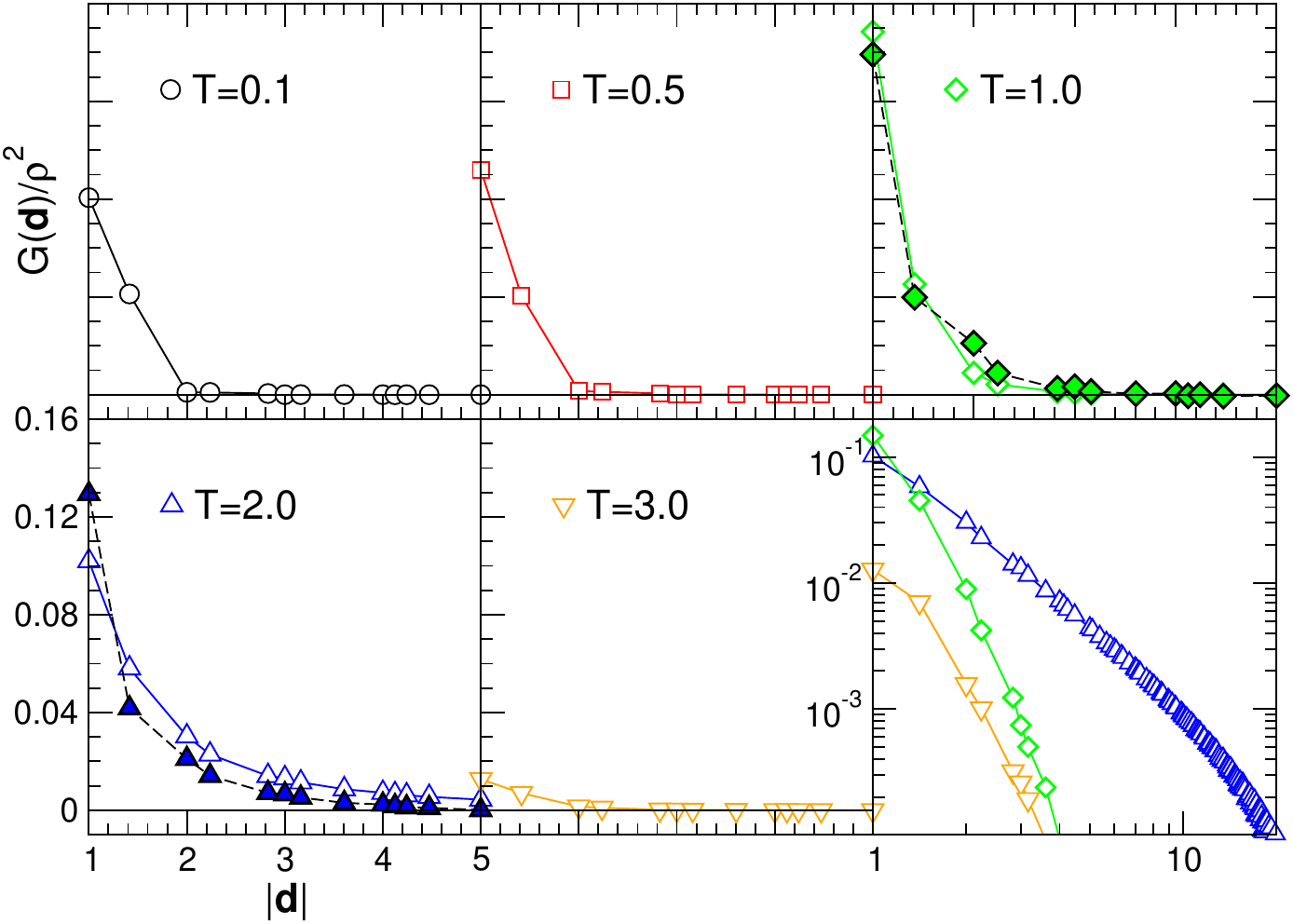}
\end{subfigure}
\begin{subfigure}[b]{0.45\linewidth}
\includegraphics[width=0.95\linewidth,clip]{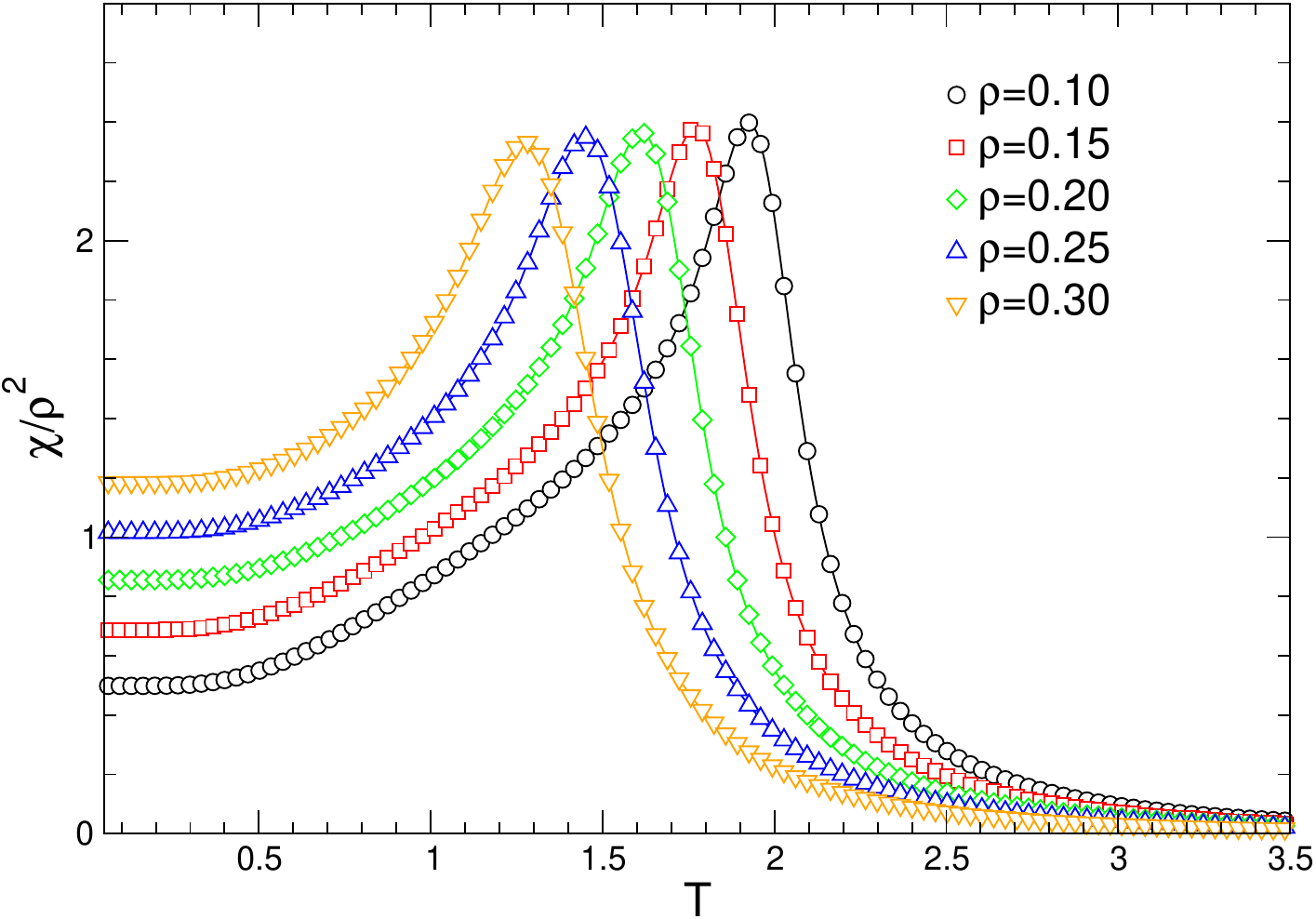}
\end{subfigure}
\caption{
(a) Correlation function as function of the distance $|\mathbf{d}|\ge 1$, for $\rhop=0.1$ and $L=50$.
The filled symbols for $T=1$ and $T=2$ are the corresponding values from the MC simulation. At
large distances, the correlation is always slightly negative except around the critical region $T=2$.
(b) Susceptibility as function of temperature for several particle densities. 
}\label{fig_corr}
\end{figure*}
%
In figure \ref{fig_corr}(a) is plotted the residual correlation function $G(|\mathbf{d}|)$, averaged
on different points of the lattices and for different temperatures, in the case $\rhop=0.1$ and
$L=50$. At low temperature $G$ is maximum near $|\mathbf{d}|=1$ and decreases rapidly with the distance,
as particles tends to cluster. $G$ is negative at larger distances but very close to zero, leading to
a slightly anti-correlation effect between particles which tend to reduce their distances.
In the paramagnetic region, $T=3$ for example, $G$ is close to zero for all distances, which implies that particles are mostly uncorrelated. In the critical region, $T=2$, $G$ has a broader distribution over larger distances and stays positive. This can be explained by noticing that spin fluctuations are correlated at large distances, which leads to an indirect correlation between particles accordingly over the same range of distances. The numerical simulations lead to results that are in agreement with the
theory, see the filled symbols
in figure \ref{fig_corr}(a) for $T=1$ and $T=2$.
The amount of correlation can be quantitatively represented in figure \ref{fig_corr}(b) where $\chi$ measures the sum of residual correlation between particles. Near the critical temperature $\chi$ is maximal, and
in the low temperature regime it decreases to a non-zero limit. We can notice that $\chi$ is always positive, which is due to the main positive contribution at small distances for all the temperature range, except when $T\rightarrow \infty$ where the residual correlations are vanishing at all distances. At high temperature, we can indeed show that $\chi\rightarrow 0$ as expected by considering the asymptotic limit of the formula \eref{eq_chi}. In this limit we have the simple relations $\rhop\simeq z/(1+z)$, $Z_{\Bk}\simeq (1+z)^2$, and $g_{\Bk}\simeq 1$, which implies that $\chi\simeq 0$. At low temperature, or $T\lesssim 0.5$, $\chi$ reaches a constant and non-zero limit, due to the fact that particles are strongly correlated at short distances. The correlation strength is in addition higher between next neighboring sites than in the spin critical regime, as seen in the inset of figure \ref{fig_corr}(a), for $T=0.5$ and $T=2$. Conversely, these correlations decay faster at longer distances due to the lack of spin fluctuations at low temperature.

We can try to evaluate analytically the asymptotic behavior of the correlation function (\ref{eq_Gd}) at
large distance, in the thermodynamical limit $L\rightarrow\infty$, and around the critical point,
in order to show that it is slightly negative. The expansion is
performed in \ref{app_2}, in the regime where $|\Bd|=d\gg 1$ and the temperature is close to 
the critical transition where $m$ is small, and for a small concentration of defects $z\ll 1$.
We define the following quantities, see also \ref{app_2}
\bb
\ca=u(1+z-u^2),\; {\mathrm{and}}\; \cb=u^2z.
\ee
In this regime, we can define the spin correlation length given by $\xi_0=\sqrt{\ca}/(2|m|)$, and
a characteristic length $\Lc=\ca/(2\cb\xi_0)=\sqrt{\ca}|m|/\cb$, which is proportional to $\Lc\sim (\xi_0\rhop)^{-1}$ for small $z$ as $\cb\sim z$ and $\rhop\simeq z\FZ(0)$ with $\FZ(0)\ne 0$. 
After applying a saddle point method to \eref{eq_Gd}, see \ref{app_2} for details, we find that the dominant behavior of the correlation function is given by
\bb\label{eq_Gd_asymp0}
G(\Bd)&\simeq -\frac{z^2m^2\xi_0}{4\pi\ca^2d}
\exp\left (-\frac{d}{\xi_0}-\frac{2d}{\Lc}\right )I_0^2\left (\frac{d}{\Lc}\right )
(1+{\mathcal O}({\xi_0}^{-2})),
\ee
with the two asymptotic regimes, depending on whether $d$ is larger or smaller than $\Lc$
\bb\nn
G(\Bd)&\simeq -\frac{z^2m^2\xi_0}{4\pi\ca^2d}
\exp\left (-\frac{d}{\xi_0}\right )(1+{\mathcal O}({\xi_0}^{-2})),\;1\ll d\ll \Lc,
\\ \label{eq_Gd_asymp}
&\simeq -\frac{z^2m^2}{16\pi^2\ca\cb d^2}
\exp\left (-\frac{d}{\xi_0}\right )(1+{\mathcal O}({\xi_0}^{-2})),\;d\gg\Lc.
\ee
The correlation function is slightly negative since $m$ is considered as small, and proportional to
$z^2$, or $\rhop^2$. This result is coherent with the data presented in figure \eref{fig_corr} as at some 
distance $G$ is negative outside the critical regime where $G$ is always positive and displays
power law behavior as all defects are correlated at long distances. 
Defects tend therefore to cluster and make the residual correlation
function negative above some typical distance that depends on the temperature.
We can notice that \eref{eq_Gd_asymp} displays a crossover behavior between
two regimes where the correlation function, apart from the exponential decay, presents 
corrections in $d^{-2}$ when $d>\Lc$, and
in $d^{-1}$ at shorter distances. The presence of two lengths $\xi_0$ and $\Lc$ is characteristic
of a system where the spin medium has its own dynamics and correlations characterized by $\xi_0$, while $\Lc$ is associated with the mediated spin correlations between defects and which should depends on $\xi_0$ as well as the concentration $\rhop$.
%
\section{Conclusion\label{sect_e}}
%
We have quantified the strength of the attractive correlations between diffusive defects in a magnetic medium, which are mediated by spin fluctuations that induce a crystallization of the defects at low temperature, namely a phenomenon of particle clustering in the low temperature region with weak spin fluctuations. We have used a fermionization technique of the action based on Grassmann variables which leads to an exact expression with non-local quartic contributions, and corrections 
are expected to be relatively small at small concentration of particles. Strong deviations are however observed when the defect concentration is large. The fermionization technique using
the Grassmann algebra is intrinsic to the Ising model and the correspondence between pairs of Grassmann
variables and particle density, such that $n_{\Br}\simeq za_{\Br}\bar a_{\Br}$, leads to a simple representation of the thermodynamical quantities of the defects.
We found in the MC simulations that the specific heat presents two peaks, one corresponding to the usual spin critical transition with a reduced temperature, and the other peak, around $T=0.5$, seems to correspond to the condensation of defects when the spin fluctuations are strongly reduced, at least for a system with periodic conditions. However we found that this phenomenon is not present in the fermionic theory, and is replaced by a macroscopic entropy at low temperature, which is nonetheless lower than the entropy for the arrangement of independent hard core particles on a square lattice. This discrepancy needs to be addressed in a future study. The correlation function between defects presents short range
attraction, as expected, and anticorrelation at larger distances. This is supported by the MC simulations
and by an asymptotic study of these correlations. Moreover, the susceptibility shows that the critical regime, for which long-range spin correlations are dominant, contributes the most to the dynamics of the defects.

\ack
This work was supported by the National Research Foundation of Korea through the Basic Science Research Program (Grant No. 2022R1A2C1012532).
%
%
\appendix
\section{\label{app_1} Corrections including the quartic terms}
The Grassmannian action $\SP$ for the Ising model in presence of particles is defined by
equation \eref{ZGC}, and we have made the approximation that only quadratic terms are dominant for
the evaluation of the thermodynamical quantities, see 
equation \eref{ZSpart}. If we take into account the quartic terms that appears as factors of the activity $z$, we have the following action instead
\bb\nn
\SP&=&\SO+z\sum_{\Br}\ar\abr\e^{-S_{\Br}}=\SO+z\sum_{\Br}\ar\abr-zu^2\sum_{\Br}
\ar\arx\abr\abry
\\ \label{Stot_Ising}
&=&\SO+z\sum_{\Br}\ar\abr+\Sint.
\ee
The quartic terms in $\Sint$ can rewritten in the Fourier space, using as before the
transformations $\ar=L^{-1}\sum_{\Bk}\ak\e^{i\Bk.\Br}$ and $\abr=L^{-1}\sum_{\Bk}\abk\e^{-i\Bk.\Br}$
\bb\nn
\Sint&=&-zu^2\frac{1}{N}\sum_{{\Bk}_i}a_{{\Bk}_1}a_{{\Bk}_2}\bar a_{{\Bk}_3}\bar a_{{\Bk}_4}
\e^{-i({\Bk}_2)_x+i({\Bk}_4)_y}\delta_{\Bk_1+\Bk_2-\Bk_3-\Bk_4,\bO}
\\
&=&-zu^2\frac{1}{N}\sum_{{\Bk}_i}a_{{\Bk}_1}a_{{\Bk}_2}\bar a_{{\Bk}_3}\bar a_{{\Bk}_4}
V_{\Bk_2,\Bk_4}\delta_{\Bk_1+\Bk_2-\Bk_3-\Bk_4,\bO},
\ee
where the Grassmannian potential $V$ is defined by
\bb
V_{\Bk_2,\Bk_4}=\e^{-i(\Bk_2)_x+i(\Bk_4)_y}.
\ee
In the manuscript, we have renamed for convenience 
the momenta $k_x\rightarrow -k_x$ and therefore we will consider
\bb
V_{\Bk_2,\Bk_4}=\e^{i(\Bk_2)_x+i(\Bk_4)_y}.
\ee
We would like to approximate the quartic term as the product of a quadratic operator and a scalar
which corresponds to some thermal average, and obtain self-consistent equations by neglecting
higher order fluctuations \cite{Clusel:2009}. The aim is to estimate the contribution magnitude of the quadratic terms in the thermodynamics quantities. There are three different possibilities to contract a pair of Grassmann variables in the quartic term, where $\Bk_1$ can be contracted with $\Bk_2$, $\Bk_3$, or $\Bk_4$. We define 
\bb
a_{\Bk_1}a_{\Bk_2}=\langle a_{\Bk_1}a_{\Bk_2}\rangle+(a_{\Bk_1}a_{\Bk_2}-\langle a_{\Bk_1}a_{\Bk_2}\rangle)
=\langle a_{\Bk_1}a_{\Bk_2}\rangle+\delta_{a_{\Bk_1}a_{\Bk_2}},
\ee
where $\delta_{a_{\Bk_1}a_{\Bk_2}}$ is the fluctuation of the pair $a_{\Bk_1}a_{\Bk_2}$.
We can expand the quartic term, using the three possible contractions, as
\bb\nn
a_{{\Bk}_1}a_{{\Bk}_2}\bar a_{{\Bk}_3}\bar a_{{\Bk}_4}&=
\frac{1}{3}\Big [
(\langle a_{\Bk_1}a_{\Bk_2}\rangle+\delta_{a_{\Bk_1}a_{\Bk_2}})
(\langle \bar a_{\Bk_3}\bar a_{\Bk_4}\rangle+\delta_{\bar a_{\Bk_3}a_{\Bk_4}})
\\ \nn
&-
(\langle a_{\Bk_1}\bar a_{\Bk_3}\rangle+\delta_{a_{\Bk_1}\bar a_{\Bk_3}})
(\langle a_{\Bk_2}\bar a_{\Bk_4}\rangle+\delta_{a_{\Bk_2}\bar a_{\Bk_4}})
\\
&+(\langle a_{\Bk_1}\bar a_{\Bk_4}\rangle+\delta_{a_{\Bk_1}\bar a_{\Bk_4}})
(\langle a_{\Bk_2}\bar a_{\Bk_3}\rangle+\delta_{a_{\Bk_2}\bar a_{\Bk_3}})
\Big ].
\ee
If we neglect the terms in $\delta^2$, we obtain 
\bb
\nn
a_{{\Bk}_1}a_{{\Bk}_2}\bar a_{{\Bk}_3}\bar a_{{\Bk}_4}\simeq 
\frac{1}{3}\Big [
-\langle a_{\Bk_1}a_{\Bk_2}\rangle \langle \bar a_{\Bk_3}\bar a_{\Bk_4}\rangle
+\langle a_{\Bk_1}\bar a_{\Bk_3}\rangle \langle a_{\Bk_2}\bar a_{\Bk_4}\rangle
-\langle a_{\Bk_2}\bar a_{\Bk_3}\rangle \langle a_{\Bk_1}\bar a_{\Bk_4}\rangle
\\ \nn
+\langle a_{\Bk_1}a_{\Bk_2}\rangle \bar a_{\Bk_3}\bar a_{\Bk_4}
+\langle \bar a_{\Bk_3}\bar a_{\Bk_4}\rangle a_{\Bk_1}a_{\Bk_2}
-\langle a_{\Bk_1}\bar a_{\Bk_3}\rangle a_{\Bk_2}\bar a_{\Bk_4}
-\langle a_{\Bk_2}\bar a_{\Bk_4}\rangle a_{\Bk_1}\bar a_{\Bk_3}
\\
+\langle a_{\Bk_1}\bar a_{\Bk_4}\rangle a_{\Bk_2}\bar a_{\Bk_3}
+\langle a_{\Bk_2}\bar a_{\Bk_3}\rangle a_{\Bk_1}\bar a_{\Bk_4}
\Big ].
\ee
The only nonzero correlators are $\langle a_{\Bk}\bar a_{\Bk}\rangle$, $\langle a_{\Bk}a_{-\Bk}\rangle$
and $\langle \bar a_{\Bk}\bar a_{-\Bk}\rangle$, since these are the only possibilities in $\SO$.
The sum over momenta gives therefore,
after grouping the terms $(a_{\Bk}\bar a_{\Bk},a_{\Bk}a_{-\Bk},\bar a_{\Bk}\bar a_{-\Bk})$
\bb
\nn\fl
\Sint=-\frac{zu^2}{3}\frac{1}{N}\sum_{\Bk,\Bk'}
\Big [ -\langle a_{\Bk}a_{-\Bk}\rangle \langle \bar a_{\Bk'}\bar a_{-\Bk'}\rangle V_{-\Bk,-\Bk'}
+\langle a_{\Bk}\bar a_{\Bk}\rangle \langle a_{\Bk'}\bar a_{\Bk'}\rangle (V_{\Bk',\Bk'}
-V_{\Bk,\Bk'})
\\ \nn\fl
+\langle a_{\Bk'}a_{-\Bk'}\rangle \bar a_{\Bk}\bar a_{-\Bk}  V_{-\Bk',-\Bk}
+\langle \bar a_{\Bk'}\bar a_{-\Bk'}\rangle a_{\Bk}a_{-\Bk} V_{-\Bk,-\Bk'}
+\langle a_{\Bk'}\bar a_{\Bk'}\rangle a_{\Bk}\bar a_{\Bk} (-V_{\Bk,\Bk}-V_{\Bk',\Bk'}
+V_{\Bk,\Bk'}+V_{\Bk',\Bk})
\Big ].
\ee
The last step is to restrict the double sum to half the Brillouin zone, see figure \ref{fig_Brillouin}, and keep the quadratic terms corresponding to
$(a_{\Bk}\bar a_{\Bk},a_{-\Bk}\bar a_{-\Bk},a_{\Bk}a_{-\Bk},\bar a_{\Bk}\bar a_{-\Bk})$
\bb
\nn\fl
\Sint=&-&\frac{zu^2}{3}\frac{1}{N}{\sum_{\Bk,\Bk'}}^{'}
\Big [ -\langle a_{\Bk}a_{-\Bk}\rangle \langle \bar a_{\Bk'}\bar a_{-\Bk'}\rangle 
(V_{-\Bk,-\Bk'}-V_{\Bk,-\Bk'}-V_{-\Bk,\Bk'}+V_{\Bk,\Bk'})
\\ \nn
&+&\langle a_{\Bk}\bar a_{\Bk}\rangle 
\Big \{
\langle a_{\Bk'}\bar a_{\Bk'}\rangle (V_{\Bk',\Bk'}-V_{\Bk,\Bk'})
+\langle a_{-\Bk'}\bar a_{-\Bk'}\rangle (V_{-\Bk',-\Bk'}
-V_{\Bk,-\Bk'})
\Big \}
\\ 
&+&\langle a_{-\Bk}\bar a_{-\Bk}\rangle
\Big \{
\langle a_{\Bk'}\bar a_{\Bk'}\rangle (V_{\Bk',\Bk'}
-V_{-\Bk,\Bk'})
+
\langle a_{-\Bk'}\bar a_{-\Bk'}\rangle (V_{-\Bk',-\Bk'}
-V_{-\Bk,-\Bk'})\Big \}\Big ]
\\ \nn
\\ \nn
&-&\frac{zu^2}{3}\frac{1}{N}{\sum_{\Bk,\Bk'}}^{'}
\Big [
a_{\Bk}a_{-\Bk}\langle \bar a_{\Bk'}\bar a_{-\Bk'}\rangle  (V_{-\Bk,-\Bk'}-V_{\Bk,-\Bk'}-V_{-\Bk,\Bk'}+V_{\Bk,\Bk'})
\\ \nn
&+&
\bar a_{\Bk}\bar a_{-\Bk}\langle a_{\Bk'}a_{-\Bk'}\rangle  (V_{-\Bk',-\Bk}-V_{\Bk',-\Bk}-V_{-\Bk',\Bk}
+V_{\Bk',\Bk})
\\ \nn
&+& a_{\Bk}\bar a_{\Bk} \Big \{
\langle a_{\Bk'}\bar a_{\Bk'}\rangle(-V_{\Bk,\Bk}-V_{\Bk',\Bk'}
+V_{\Bk,\Bk'}+V_{\Bk',\Bk})
\\ \nn
&+&\langle a_{-\Bk'}\bar a_{-\Bk'}\rangle(-V_{\Bk,\Bk}-V_{-\Bk',-\Bk'}
+V_{\Bk,-\Bk'}+V_{-\Bk',\Bk})
\Big \}
\\ \nn
&+& a_{-\Bk}\bar a_{-\Bk} \Big \{
\langle a_{\Bk'}\bar a_{\Bk'}\rangle(-V_{-\Bk,-\Bk}-V_{\Bk',\Bk'}
+V_{-\Bk,\Bk'}+V_{\Bk',-\Bk})
\\ 
&+&\langle a_{-\Bk'}\bar a_{-\Bk'}\rangle(-V_{-\Bk,-\Bk}-V_{-\Bk',-\Bk'}
+V_{-\Bk,-\Bk'}+V_{-\Bk',-\Bk})
\Big \}
\Big ].
\ee
This yields, after replacing the $V_{\Bk,\Bk'}$s by their value
\bb
\nn
\Sint=-\frac{zu^2}{3}\frac{1}{N}{\sum_{\Bk,\Bk'}}^{'}
\Big [ 4\sin(k_x)\sin(k_y')\langle a_{\Bk}a_{-\Bk}\rangle \langle \bar a_{\Bk'}\bar a_{-\Bk'}\rangle 
\\ \nn
+\e^{ik_y'}(\e^{ik_x'}-\e^{ik_x})\langle a_{\Bk}\bar a_{\Bk}\rangle \langle a_{\Bk'}\bar a_{\Bk'}\rangle
\\ \nn
+(\e^{ik_x}-\e^{-ik_x'})(\e^{ik_y}-\e^{-ik_y'})\langle a_{\Bk}\bar a_{\Bk}\rangle \langle a_{-\Bk'}\bar a_{-\Bk'}\rangle
\\ \nn
+\e^{-ik_y'}(\e^{-ik_x'}-\e^{-ik_x})\langle a_{-\Bk}\bar a_{-\Bk}\rangle \langle a_{-\Bk'}\bar a_{-\Bk'}\rangle \Big ]
\\ \nn
\\ \nn
+\frac{zu^2}{3}\frac{1}{N}{\sum_{\Bk,\Bk'}}^{'}
\Big [
4a_{\Bk}a_{-\Bk}\langle \bar a_{\Bk'}\bar a_{-\Bk'}\rangle  \sin(k_x)\sin(k_y') 
+4\bar a_{\Bk}\bar a_{-\Bk}\langle a_{\Bk'}a_{-\Bk'}\rangle \sin(k_y) \sin(k_x')
\\ \nn
+a_{\Bk}\bar a_{\Bk} \Big \{
\langle a_{\Bk'}\bar a_{\Bk'}\rangle (\e^{ik_x}-\e^{ik_x'})(\e^{ik_y}-\e^{ik_y'})
+\langle a_{-\Bk'}\bar a_{-\Bk'}\rangle (\e^{ik_x}-\e^{-ik_x'})(\e^{ik_y}-\e^{-ik_y'})
\Big \}
\\ \nn
+a_{-\Bk}\bar a_{-\Bk} \Big \{
\langle a_{\Bk'}\bar a_{\Bk'}\rangle (\e^{-ik_x}-\e^{ik_x'})(\e^{-ik_y}-\e^{ik_y'})
+\langle a_{-\Bk'}\bar a_{-\Bk'}\rangle (\e^{-ik_x}-\e^{-ik_x'})(\e^{-ik_y}-\e^{-ik_y'})
\Big \}
\Big ].
\ee
Let define the following averages, all real quantities
\bb
\vx=\frac{2i}{N}{\sum_{\Bk}}^{'}\langle \bar a_{\Bk}\bar a_{-\Bk}\rangle \sin(k_y),\;
\vy=-\frac{2i}{N}{\sum_{\Bk}}^{'}\langle a_{\Bk}a_{-\Bk}\rangle \sin(k_x),
\ee
and
\bb\nn
\uuxy=\frac{1}{N}{\sum_{\Bk}}^{'}\langle a_{\Bk}\bar a_{\Bk}\rangle+
\langle a_{-\Bk}\bar a_{-\Bk}\rangle,\;
\\ \nn
\uuy=\frac{1}{N}{\sum_{\Bk}}^{'}\langle a_{\Bk}\bar a_{\Bk}\rangle \e^{ik_x}+
\langle a_{-\Bk}\bar a_{-\Bk}\rangle \e^{-ik_x},
\\ \nn
\uux=\frac{1}{N}{\sum_{\Bk}}^{'}\langle a_{\Bk}\bar a_{\Bk}\rangle \e^{ik_y}+
\langle a_{-\Bk}\bar a_{-\Bk}\rangle \e^{-ik_y}, \;
\\ 
\uu=\frac{1}{N}{\sum_{\Bk}}^{'}\langle a_{\Bk}\bar a_{\Bk}\rangle \e^{ik_x+ik_y}+
\langle a_{-\Bk}\bar a_{-\Bk}\rangle \e^{-ik_x-ik_y}.
\ee
The modified action for the quartic term is then equal to
\bb
\nn\fl
\Sint&=-\frac{zu^2}{3}N\Big (\vx\vy+\uu\uuxy-\uux\uuy \Big )
\\
&+\frac{zu^2}{3}{\sum_{\Bk}}^{'}
\Big [-2i\sin(k_x)\vx a_{\Bk}a_{-\Bk} +2i\sin(k_y)\vy\bar a_{\Bk}\bar a_{-\Bk}
\\ \nn\fl
&+ a_{\Bk}\bar a_{\Bk} \Big \{
\uuxy \e^{ik_x+ik_y}-\uux \e^{ik_x}-\uuy \e^{ik_y}+\uu
\Big \}
\\ \nn\fl
&+ a_{-\Bk}\bar a_{-\Bk} \Big \{
\uuxy \e^{-ik_x-ik_y}-\uux \e^{-ik_x}-\uuy \e^{-ik_y}+\uu
\Big \}
\Big ]
\\ \fl\nn
&=-\frac{zu^2}{3}N\Big (\vx\vy+\uu\uuxy-\uux\uuy \Big )
\\ \fl
&+\frac{zu^2}{3}{\sum_{\Bk}}^{'}
\Big [-2i\sin(k_x)\vx a_{\Bk}a_{-\Bk} +2i\sin(k_y)\vy\bar a_{\Bk}\bar a_{-\Bk}
+ h_{\Bk}a_{\Bk}\bar a_{\Bk}
+ \bar h_{\Bk}a_{-\Bk}\bar a_{-\Bk} 
\Big ].
\ee
with
\bb\nn
h_{\Bk}=\uuxy \e^{ik_x+ik_y}-\uux \e^{ik_x}-\uuy \e^{ik_y}+\uu,
\\
\bar h_{\Bk}=h_{-\Bk}=\uuxy \e^{-ik_x-ik_y}-\uux \e^{-ik_x}-\uuy \e^{-ik_y}+\uu.
\ee
Therefore we can write the total action containing the Ising part and interacting part 
with particles as 
\bb\nn
\SP&=-\frac{zu^2}{3}N\Big (\vx\vy+\uu\uuxy-\uux\uuy \Big )
\\ \nn 
&+{\sum_{\Bk}}^{'}\Big [
(z+g_{\Bk}+\frac{zu^2}{3}h_{\Bk})\ak\abk+(z+\bar g_{\Bk}+\frac{zu^2}{3}\bar h_{\Bk})\amk\abmk
\\
&-2iu\sin(k_x)(1+\frac{zu}{3}\vx)\ak\amk+2iu\sin(k_y)(1+\frac{zu}{3}\vy)\abk\abmk
\Big ].
\ee
The single terms $z$ in the coefficients of $\ak\abk$ and $\amk\abmk$ come from $z\sum_{\Br}\ar\abr$ in equation \eref{Stot_Ising} and which can be decomposed as $z{\sum_{\Bk}}^{'}(\ak\abk+\amk\abmk)$. 
We can therefore write the total partition function in the Grand Canonical Ensemble as
\bb\fl
\ZSP=2^{N}\cosh(\beta J)^{2N}{\prod_{\Bk}}^{'}\int d\bar a_{\Bk}da_{\Bk}d\bar a_{-\Bk}da_{-\Bk}
\exp(\SP)
\\ \fl \nn
=2^{N}\cosh(\beta J)^{2N}\exp\left [-zu^2N/3\Big (\vx\vy+\uu\uuxy-\uux\uuy \Big )\right ]
\\ \nn\fl
\times{\prod_{\Bk}}^{'}
\Big [
(z+g_{\Bk}+\frac{zu^2}{3}h_{\Bk})(z+\bar g_{\Bk}+\frac{zu^2}{3}\bar h_{\Bk})
-4u^2\sin(k_x)\sin(k_y)(1+\frac{zu}{3}\vx)(1+\frac{zu}{3}\vy)
\Big ].
\ee
In the following, we will set
%
\bb\fl\fl
Z_{\Bk}=(z+g_{\Bk}+\frac{zu^2}{3}h_{\Bk})(z+\bar g_{\Bk}+\frac{zu^2}{3}\bar h_{\Bk})
-4u^2\sin(k_x)\sin(k_y)(1+\frac{zu}{3}\vx)(1+\frac{zu}{3}\vy),
\ee
such that $\ZSP$ is proportional to ${\prod_{\Bk}}^{'}Z_{\Bk}$. The different averages can be performed after integration
\bb\nn
\vx=-\frac{4u}{N}{\sum_{\Bk}}^{'}\frac{1}{Z_{\Bk}}\sin(k_x)\sin(k_y)(1+\frac{zu}{3}\vx),
\\ \label{eq_vs}
\vy=-\frac{4u}{N}{\sum_{\Bk}}^{'}\frac{1}{Z_{\Bk}}\sin(k_x)\sin(k_y)(1+\frac{zu}{3}\vy),
\ee
and
\bb\nn
\uuxy
=\frac{2}{N}\Re{\sum_{\Bk}}^{'}\frac{1}{Z_{\Bk}} 
(z+g_{\Bk}+\frac{zu^2}{3}h_{\Bk}),
\\ \nn
\uux=\frac{2}{N}\Re{\sum_{\Bk}}^{'}\frac{1}{Z_{\Bk}} 
(z+g_{\Bk}+\frac{zu^2}{3}h_{\Bk})\e^{-ik_y},
\\ \nn
\uuy=\frac{2}{N}\Re{\sum_{\Bk}}^{'}\frac{1}{Z_{\Bk}} 
(z+g_{\Bk}+\frac{zu^2}{3}h_{\Bk})\e^{-ik_x},
\\ \label{eq_us}
\uu=\frac{2}{N}\Re{\sum_{\Bk}}^{'}\frac{1}{Z_{\Bk}} 
(z+g_{\Bk}+\frac{zu^2}{3}h_{\Bk})\e^{-ik_x-ik_y}.
\ee
%
From equation \eref{eq_vs}, by space invariance in both directions, we have $\vx=\vy$ as a solution. Therefore we are left with real five quantities to compute
self-consistently: $(\vx,\uu,\uux,\uuy,\uuxy)$. Function $\FZ(z)$ is then defined by $\FZ(z)=\partial_z \log(\ZSP)$, or
\bb
\FZ(z)=\frac{\partial}{\partial z}\Big [
-\frac{zu^2}{3}\Big (\vx\vy+\uu\uuxy-\uux\uuy \Big )\Big ]+
\frac{\partial}{\partial z}\frac{1}{N}{\sum_{\Bk}}^{'}\log Z_k.
\ee
If we neglect the quartic term, we recover instead the original function
\bb\nn
\FZ(z)&=
\frac{\partial}{\partial z}\frac{1}{N}{\sum_{\Bk}}^{'}\log \Big [
(z+g_{\Bk})(z+\bar g_{\Bk})-4u^2\sin(k_x)\sin(k_y)\Big ]
\\ 
&=\frac{1}{N}{\sum_{\Bk}}^{'}\frac{2z+g_{\Bk}+\bar g_{\Bk}}{
(z+g_{\Bk})(z+\bar g_{\Bk})-4u^2\sin(k_x)\sin(k_y)}.
\ee
%
%
\section{\label{app_2}Two-point correlation functions and their asymptotic behavior}
We consider in this appendix the expression given by \eref{eq_Gd}, and try to obtain
some analytical limit when the concentration is low, or $z\ll 1$, and the distance
$|\Bd|$ large. We begin to rewrite the Boltzmann weight $Z_{\Bk}$ as

\bb\fl\nn
Z_{\Bk}=(1+z)^2+u^2(2+u^2)-2u(1+z-u^2)(\cos(k_x)+\cos(k_y))-2u^2z\cos(k_x+k_y)
\\
=m^2+2\ca(2-\cos(k_x)-\cos(k_y))+2\cb(1-\cos(k_x+k_y)),
\ee
where
\bb
m^2=(u^2+2u-1-z)^2,\; \ca=u(1+z-u^2),\; {\mathrm{and}}\; \cb=u^2z.
\ee
In order to simplify the expression \eref{eq_Gd},
we would like to take the limit $L\rightarrow\infty$ and replace the momenta
$k_x$ and $k_y$ by their continuous counterparts, while keeping $d_x$ and $d_y$ as integers with $\Bd\ne\bO$. We will need to introduce the following quantities
\bb\label{eq_I}
I_{\epsilon\epsilon'}^{\sigma}(\Bd)
=\int_{-\pi}^{\pi}\frac{dk_x}{2\pi}\int_{0}^{\pi}\frac{dk_y}{2\pi}
\frac{\e^{i\epsilon k_x+i\epsilon'k_y+i\sigma(k_xd_x+k_yd_y)}}{Z_{\Bk}},
\ee
where $\sigma=\pm 1$, and $\epsilon$, $\epsilon'$ take the values $1$, $-1$, or $0$.
It is clear that $\bar I_{\epsilon\epsilon'}^{\sigma}=I_{-\epsilon-\epsilon'}^{-\sigma}$.
The correlation function (\ref{eq_Gd}) can be expressed using these functions, and we find
\bb\nn
G(\Bd)=-z^2u^2\left ( I_{10}^{1}+I_{-10}^{-1}-I_{10}^{-1}-I_{-10}^{1} \right )
\left ( I_{01}^{1}+I_{0-1}^{-1}-I_{0-1}^{1}-I_{01}^{-1} \right )
\\ \nn
-z^2\left [ (1+z)(I_{00}^{1}+I_{00}^{-1})-u(I_{10}^{1}+I_{-10}^{-1}+I_{01}^{1}+I_{0-1}^{-1})
-u^2(I_{11}^{1}+I_{-1-1}^{-1})\right ]
\\ \label{eq_GdI}
\times \left [ (1+z)(I_{00}^{1}+I_{00}^{-1})-u(I_{10}^{-1}+I_{-10}^{1}+I_{01}^{-1}+I_{0-1}^{1})
-u^2(I_{11}^{-1}+I_{-1-1}^{1})\right ].
\ee
In this expression, only the real part of $I_{\epsilon\epsilon'}^{\sigma}$ is needed, as 
each term is added to its conjugate. To evaluate $I_{\epsilon\epsilon'}^{\sigma}$, it is 
useful to separate the variables $k_x$ and $k_y$, using the following transformation
\bb\nn
I_{\epsilon\epsilon'}^{\sigma}(\Bd)
=\int_0^{\infty}d\lambda \e^{-\lambda(m^2+4\ca+2\cb)}
\int_{-\pi}^{\pi}\frac{dk_x}{2\pi}\int_{0}^{\pi}\frac{dk_y}{2\pi}
\e^{i\epsilon k_x+i\epsilon'k_y+i\sigma(k_xd_x+k_yd_y)}
\\ 
\times \exp\Big [2\ca\lambda(\cos(k_x)+\cos(k_y))
+2\cb\lambda\cos(k_x+k_y)\Big ].
\ee
We can then use the series representation of the cosine exponentials in terms of modified Bessel functions, in order to integrate over the momenta separately
\bb\nn
I_{\epsilon\epsilon'}^{\sigma}(\Bd)
=\int_0^{\infty}d\lambda \e^{-\lambda(m^2+4\ca+2\cb)}
\int_{-\pi}^{\pi}\frac{dk_x}{2\pi}\int_{0}^{\pi}\frac{dk_y}{2\pi}
\e^{i\epsilon k_x+i\epsilon'k_y+i\sigma(k_xd_x+k_yd_y)}
\\ \label{eq_I2}
\times \sum_{n,n'n''}I_{n}(2\ca\lambda)I_{n'}(2\ca\lambda)I_{n''}(2\cb\lambda)
\e^{ink_x+in'k_y+in''(k_x+k_y)},
\ee
where the three indices run over the positive and negative integers. After integrating over 
the momenta, we obtain the real part as the following series
\bb\fl\label{eq_I3}
\Re\left (I_{\epsilon\epsilon'}^{\sigma}(\Bd)\right )=
\frac{1}{2}\int_0^{\infty}d\lambda \e^{-\lambda(m^2+4\ca+2\cb)}
\sum_{n}
I_{n+\epsilon+\sigma d_x}(2\ca\lambda)I_{n+\epsilon'+\sigma d_y}(2\ca\lambda)
I_{n}(2\cb\lambda).
\ee
This expression is exact in the limit $L\rightarrow\infty$.
To simplify the problem, we will consider the small concentration limit, where $z\ll 1$. Since 
$G(\Bd)$ in \eref{eq_GdI} is proportional to $z^2$, we can take the term $n=0$ in the
previous expression since $I_n(2\cb\lambda)\simeq  (\cb\lambda)^{|n|}/\Gamma(1+|n|)$ for small argument
and only the term $n=0$ is non negligible. We then have the approximation
%
\bb\label{eq_Iapprox}\fl
\Re\left (I_{\epsilon\epsilon'}^{\sigma}(\Bd)\right )\simeq
\frac{1}{2}\int_0^{\infty}d\lambda \e^{-\lambda(m^2+4\ca+2\cb)}
I_{\epsilon+\sigma d_x}(2\ca\lambda)I_{\epsilon'+\sigma d_y}(2\ca\lambda)I_0(2\cb\lambda).
\ee
We would like to take the asymptotic limit of the previous, when $|\Bd|$ is large, in
order to check in particular the sign of the correlation function at large distances.
We can use the following asymptotic expansion of the modified Bessel function for large order
\cite{book:Abramowitz}
\bb\label{eq_BesselApprox}
I_{d}(dz)\simeq \frac{1}{\sqrt{2\pi d}}\frac{\e^{d\eta(z)}}{(1+z^2)^{1/4}},
\;\eta(z)=\sqrt{1+z^2}+\log\left (\frac{z}{1+\sqrt{1+z^2}}\right ).
\ee
Then \eref{eq_Iapprox} can be rewritten as
\bb\fl
\Re\left (I_{\epsilon\epsilon'}^{\sigma}(\Bd)\right )\simeq
\frac{1}{4\pi\sqrt{d_1d_2}}\int_0^{\infty}d\lambda 
\frac{\e^{-\lambda (m^2+4\ca+2\cb)+d_1\eta(2\ca\lambda/d_1)+d_2\eta(2\ca\lambda/d_2)}}
{(1+(2\ca\lambda/d_1)^2)^{1/4}(1+(2\ca\lambda/d_2)^2)^{1/4}}
I_0(2\cb\lambda),
\ee
where $d_1=|d_x+\epsilon\sigma|$ and $d_2=|d_y+\epsilon'\sigma|$.
We can then use the saddle point approximation to the previous integral, in the limit where
$d_1$ and $d_2$ are large. We define the following ratio $r=d_1/d_2$, and perform a change
of variable $v=2\ca\lambda/d_1$ such that
%
%
\bb\fl \label{eq_Iepsilon}
\Re\left (I_{\epsilon\epsilon'}^{\sigma}(\Bd)\right )\simeq
\frac{\sqrt{r}}{8\pi\ca}\int_0^{\infty}dv
\frac{\e^{-2d_1v(\mO^2+\cb/(2\ca)) +d_1\eta(v)+d_1\eta(rv)/r}}
{(1+v^2)^{1/4}(1+r^2v^2)^{1/4}}I_0\left(\frac{vd_1\cb}{\ca}\right),
\ee
where 
\bb
\mO^2=\frac{m^2}{4\ca}+1.
\ee
The integral is well defined if $\mO^2>1$. Since $d_1$ is large, we can a priori expand the Bessel function $I_0(z)\simeq e^z/\sqrt{2\pi z}$, and simplify the previous expression
\bb
\Re\left (I_{\epsilon\epsilon'}^{\sigma}(\Bd)\right )\simeq
\frac{\sqrt{r}}{4(2\pi)^{3/2}\sqrt{\ca\cb d_1}}\int_0^{\infty}dv
\frac{\e^{-2d_1v\mO^2 +d_1\eta(v)+d_1\eta(rv)/r}}
{\sqrt{v}(1+v^2)^{1/4}(1+r^2v^2)^{1/4}}.
\ee
This expression should be valid if the distance $d_1$ is larger than some characteristic length related to the inter-particle distance ($\rho^{-1/2}\propto z^{-1/2}$ for $z$ small) and a typical length defined
by the saddle point solution $v^*$, which, from the argument of the exponential, is homogeneous to a length, as the product $d_1v^*z$ or $d_1v^*\rho$ should be dimensionless. Here a typical length would be $(v^*\rho)^{-1}$.
This translates into saying that $d_1\gg \ca(v^*\cb)^{-1}$ in the argument of $I_0$ above in equation
\eref{eq_Iepsilon}. 
Otherwise, we should treat the product $\e^{-vd_1\cb/\ca}I_0\left(vd_1\cb/\ca\right)$ as a regular function when we apply the saddle point method to the exponential argument which is proportional to $d_1$. In this case we should consider instead
\bb\fl \label{eq_Iepsilon2}
\Re\left (I_{\epsilon\epsilon'}^{\sigma}(\Bd)\right )\simeq
\frac{\sqrt{r}}{8\pi\ca}\int_0^{\infty}dv
\frac{\e^{-2d_1v\mO^2+d_1\eta(v)+d_1\eta(rv)/r}}{(1+v^2)^{1/4}(1+r^2v^2)^{1/4}}
\e^{-vd_1\cb/\ca}I_0\left(\frac{vd_1\cb}{\ca}\right),
\ee
and consider the extrema $v^*$ of the following function
\bb
\phi(v)=-2v\mO^2+\eta(v)+\eta(rv)/r.
\ee
Using $\eta'(v)=\sqrt{1+v^2}/v$, and $\eta''(v)=-v^{-2}/\sqrt{1+v^2}$, the saddle point
solution of $\phi'(v)=0$ is equal to the positive value
\bb\label{eq_v}
v^*=\frac{1}{2r\mO^2}\frac{\sqrt{(1+r^2)\mO^2+\sqrt{(1+r^2)^2+4r^2(\mO^4-1)}}}{\sqrt{\mO^4-1}}.
\ee
The second derivative is always negative, and we can perform a Gaussian integration after
expanding $\phi(v)\simeq\phi(v^*)+\ff(v-v^*)^2\phi''(v^*)$ up to the second order.
We obtain
\bb\fl \label{eq_saddle0}
\Re\left (I_{\epsilon\epsilon'}^{\sigma}(\Bd)\right )\simeq
\frac{\sqrt{r}}{4\ca\sqrt{-2\pi d_1\phi''(v^*)}}
\frac{\e^{d_1\phi(v^*)}\e^{-v^*d_1\cb/\ca}I_0\left(v^*d_1\cb/\ca\right)}{(1+{v^*}^2)^{1/4}(1+r^2{v^*}^2)^{1/4}}.
\ee
Near the critical temperature where $m\gtrsim 0$ or $m_0\gtrsim 1$, we can approximate the saddle solution
(\ref{eq_v}) by
\bb
v^*\simeq  \frac{\sqrt{1+r^2}}{r}\frac{\sqrt{2}}{4\sqrt{m_0-1}}= \frac{\sqrt{1+r^2}}{r}2\xi_0,
\ee
where we define $\xi_0$ as the correlation length. In this limit
we can show that $\phi(v^*)\simeq -(2\xi_0)^{-1}\sqrt{1+r^2}/r$, $\phi''(v^*)\simeq
-16\sqrt{2}(r/\sqrt{1+r^2})(m_0-1)^{3/2}=-(2\xi_0)^{-3}r/\sqrt{1+r^2}$, and after some algebra, assuming that $v^*$ is large compared to unity
\bb\fl \label{eq_saddle}
\Re\left (I_{\epsilon\epsilon'}^{\sigma}(\Bd)\right )\simeq
\frac{\sqrt{\xi_0}}{4\ca\sqrt{\pi d_{\epsilon\sigma,\epsilon'\sigma}}}
\exp\left (-\frac{d_{\epsilon\sigma,\epsilon'\sigma}}{2\xi_0}-\frac{d_{\epsilon\sigma,\epsilon'\sigma}}{\Lc}\right )
I_0\left(\frac{d_{\epsilon\sigma,\epsilon'\sigma}}{\Lc}\right),
\ee
where $d_{\epsilon\sigma,\epsilon'\sigma}=\sqrt{(d_x+\epsilon\sigma)^2+(d_y+\epsilon'\sigma)^2}
\simeq d+\sigma(\epsilon d_x+\epsilon'd_y)/d$, with $d=|\Bd|$, and $\Lc=\ca/(2\cb\xi_0)$ a characteristic length depending on the activity $z$ and spin correlation length $\xi_0$. There are two limits for \eref{eq_saddle},
depending on whether $d_{\epsilon\sigma,\epsilon'\sigma}$ is larger or smaller than $\Lc$
\bb\nn
\Re\left (I_{\epsilon\epsilon'}^{\sigma}(\Bd)\right )\simeq
\frac{\sqrt{\xi_0}}{4\ca\sqrt{\pi d_{\epsilon\sigma,\epsilon'\sigma}}}
\exp\left (-\frac{d_{\epsilon\sigma,\epsilon'\sigma}}{2\xi_0}\right ),\;d_{\epsilon\sigma,\epsilon'\sigma}<\Lc,
\\ \label{eq_saddle_limits}
\Re\left (I_{\epsilon\epsilon'}^{\sigma}(\Bd)\right )\simeq
\frac{1}{8\pi\sqrt{\ca\cb} d_{\epsilon\sigma,\epsilon'\sigma}}
\exp\left (-\frac{d_{\epsilon\sigma,\epsilon'\sigma}}{2\xi_0}\right ),\;d_{\epsilon\sigma,\epsilon'\sigma}>\Lc.
\ee
We can therefore write in general, after expanding $d_{\epsilon\sigma,\epsilon'\sigma}$ with $d$
and discarding the terms in $1/d$
\bb\label{eq_Id}
\Re\left (I_{\epsilon\epsilon'}^{\sigma}(\Bd)\right )\simeq
I(d)\exp\left (-\sigma\frac{\epsilon d_x+\epsilon' d_y}{2\xi_0 d}\right ),
\ee
with 
\bb\nn
I(d)\simeq\frac{\sqrt{\xi_0}}{4\ca\sqrt{\pi d}}
\exp\left (-\frac{d}{2\xi_0}\right ),\;d\ll\Lc,
\\ \label{eq_Idlimit}
I(d)\simeq
\frac{1}{8\pi\sqrt{\ca\cb} d}
\exp\left (-\frac{d}{2\xi_0}\right ),\;d\gg\Lc.
\ee
We can therefore deduce the form of the correlation function $G(\Bd)$ at long
distance, by replacing $I_{\epsilon\epsilon'}^{\sigma}$ in \eref{eq_GdI} by their value above,
and expand up to second order in $\xi_0^{-1}$
\bb\label{eq_Gd_app}
G(\Bd)\simeq -4z^2I(d)^2\left [m^2-\frac{u}{4{\xi_0}^2}\left (1+z-u^2
+zu\left (\frac{d_x+d_y}{d}\right )^2\right )+{\mathcal{O}}(\xi_0^{-4}) \right ].
\ee
%

%
\section*{References}
\bibliography{biblio}
\end{document}